\documentclass[aps,prd,twocolumn,preprintnumbers,nofootinbib,showpacs]{revtex4}
\usepackage{amsmath,amsfonts,amssymb,bm}
\usepackage{graphicx}
\usepackage{color}
\usepackage{subfigure}
\usepackage{multirow}
\usepackage{textcomp}
\usepackage{slashed}
\usepackage{enumitem}

\definecolor{purple}{rgb}{0.5,0,0.5}
\definecolor{blue}{rgb}{0.0,0,0.9}

\usepackage[colorlinks=true, pdfstartview=FitV, linkcolor=purple, citecolor= purple, urlcolor=blue]{hyperref}

\begin{document}


\title{Running Coupling and Running Quark Mass Effects on the Elastic Form Factors of Nucleons}




\author{Bao-Yi Yang, Hui-Hua Zhong, Muyang Chen}
\email{muyang@hunnu.edu.cn}
\affiliation{Department of Physics, Hunan Normal University, and Key Laboratory of Low-Dimensional Quantum Structures and Quantum Control of Ministry of Education, Changsha 410081, China}

\date{\today}

\begin{abstract}
 We study the elastic electric and magnetic form factors of the proton, neutron and the charged roper resonance ($G_E^p$, $G_M^p$, $G_E^n$, $G_M^n$, $G_E^R$ and $G_M^R$) systematically in a constituent quark model. Three ingredients are crucial in this study: i) the mixing between the pure S-wave and other components which produces a nonzero neutron electric form factor. ii) a running coupling constant that soften the form factors. iii) the running quark mass function, $M_q(p^2)$, which is responsible for the decreasing of the $\mu_p G_E^p(Q^2)/G_M^p(Q^2)$ as $Q^2$ increases. The produced elastic form factors of the proton and neutron match the corresponding observed values fairly well. Our study shows that $\mu_p G_E^p(Q^2)/G_M^p(Q^2) \approx M_q(Q^2/9)/M_q(0)$ upto $Q^2 \approx 4 \text{ GeV}^2$. We give predictions on the elastic form factors of the roper resonance, the electric charge and the magnetic momentum radius ratios of the roper resonance to the proton are $r^R_{E}/r^p_{E} \approx r^R_{M}/r^p_{M} \approx 1.5$.
\end{abstract}

\maketitle


\section{Introduction}\label{sec:introduction}

Proton and neutron are fundamental constituents of the universe. Though the proton was discovered more than one hundred years ago and the neutron a little later, there are still many mysteries yet to be solved. For example, the proton size puzzle \cite{Pohl2010}, the spectrum and the structure of the nucleons are not well understood yet \cite{Thiel2022,Ablikim2021,Ablikim2021a}.

In this paper we focus on the elastic form factor of the nucleons. The elastic form factor of nucleson was proposed in the middle of the 20 th century when it was found that the proton is not a point particle \cite{Chambers1956,Hofstadter1956}. The elastic electric and magnetic form factors of the proton and neutron have been measured extensively since 1960s \cite{Janssens1966,Berger1971,Bartel1973,Rock1982,Arnold1986,Lung1993,Andivahis1994,Walker1994,Milbrath1998,Jones2000,Gayou2002,Punjabi2005,Qattan2005,Geis2008,Paolone2010,Puckett2010,Zhan2011}. The Rosenbluth measurement indicates that $\mu_p G_E^p/G_M^p \approx 1$ \cite{Qattan2005}, while the polarization measurement shows that the ratio decreases dramatically as $Q^2$ increases \cite{Puckett2010}. It is suggested that this discrepancy is due to ingnoring the two photon exchange process. See the references in Ref. \cite{Pacetti2015} for this problem.

It's difficult to calculate the form factors theoretically because of the unsolvable of the nonperturbative Quantum Chromodynamics (QCD). However there are many studies on the form factors via QCD based theories or phenomenologically \cite{Leinweber1991,Hellstern1997,Baldini2006,Cloet2009,Eichmann2011,Lomon2012,Green2014,Lin2022}.
The constituent quark model, even its nonrelativistic version, describes the light baryon spectrum and decays reasonably well \cite{Horgan1973,Horgan1974,Capstick1986,Capstick2000}. Herein we study the elastic electric and magnetic form factors of the proton, neutron and the charged roper resonance systematically in a constituent quark model. In order to produce proton and neutron elastic form factors consistent with the observed values, we summarise the experience of some previous studies, and integrate the following three ingredients in our study:
\begin{enumerate}[itemindent=0em,label=\roman*)]
 \item In order to produce a nonzero neutron electric form factor, the mixing between the pure S-wave and other components is important \cite{Isgur1978}, herein we use Eq. (\ref{eq:mixingPN}) as the proton/neutron state.
 \item References \cite{Lakhina2006,Sun2023} show that a running coupling constant has a significant effect on the dynamic properties of hadrons such as form factor and decay constant. We will explore its effect on the nucleons' elastic form factors carefully.
 \item Traditional quark model studies predict that $\mu_p G_E^p/G_M^p \approx 1$ \cite{Giannini1991}, which is not consistent with the more undisputed polarization measurement. Herein we propose a running quark mass, which is responsible for the decreasing of the $\mu_p G_E^p(Q^2)/G_M^p(Q^2)$ as $Q^2$ increases. Introducing a running quark mass leads to the factorization formula of the magnetic form factor, Eq. (\ref{eq:magneticFFMQ}).
\end{enumerate}

This paper is organized as following. Section \ref{sec:modelMethod} introduces the model we use and the method to solve this model. Subsection \ref{subsec:model} introduces the interaction model, subsection \ref{subsec:wavefunction} briefly introduces the wave function of the baryon, subsection \ref{subsec:formfactor} introduces the mixing state and the formulae for the elastic form factors. Section \ref{sec:results} explains our results orderly. Subsection \ref{subsec:spectrum} discusses the nucleon and $\Delta$ spectrum shortly, subsection \ref{subsec:GEpGEn} explains the electric form factors of the proton and neutron, subsection \ref{subsec:GMpGMn} explains the magnetic form factors of the proton and neutron, subsection \ref{subsec:GERGMR} shows our predictions for the electric and magnetic form factors of the charged roper resonance. Section \ref{sec:conclusion} gives a summary and draws our conclusion. We explicate the baryon wave function completely in Appendix A. The formulae for the matrix elements of the hamiltonian are listed in Appendix B.

\section{The Model and Method}\label{sec:modelMethod}

\subsection{The interaction model}\label{subsec:model}

The framework herein is totally consistent with those for mesons \cite{Sun2023}. The mass and wave function are got by solving the Schr$\ddot{\text{o}}$dinger equation,
\begin{equation}\label{eq:schrodinger}
 H |\psi\rangle = E |\psi\rangle,
\end{equation}
where
\begin{equation}\label{eq:hamiltonian}
 H = \sum\limits_{i} T_i + \frac{1}{2}\sum\limits_{i,j} V_{ij} + C_0
\end{equation}
is the hamiltonian. The quark index $\{i,j\} = \{1,2,3 \}$. $T_i$ is the kinematic energy of the $i$-th quark, $V_{ij}$ is the potential energy between the $i$-th and $j$-th quark. $C_0$ is a constant. The baryon mass $M = E + \sum\limits_{i=1}^3 m_i$, where $m_i$ is the constituent quark mass.

The potential could be decomposed into
\begin{equation}\label{eq:interaction}
 V_{ij} = H^{\text{SI}}_{ij} + H^{\text{SS}}_{ij} + H^{\text{T}}_{ij} + H^{\text{SO}}_{ij}.
\end{equation}
$H^{\text{SI}}_{ij}$ is the spin independent potential, which is composed of a coulombic and a linear part,
\begin{equation}\label{eq:interactionSI}
 H^{\text{SI}}_{ij} = -\frac{2\alpha_s}{3r_{ij}} + \frac{b}{2}r_{ij},
\end{equation}
where $b$ is a constant and $\alpha_s$ is the coupling constant of the strong interaction. In this paper a bold character stands for a 3-dimension vector, for example, the position vector $\bm{r} = \vec{r}$. $\bm{r}_{ij} = \bm{r}_i - \bm{r}_j$ is the relative position and $r_{ij} = |\bm{r}_{ij}|$ is the distance between the $i$-th and $j$-th quark. The other three terms are spin dependent.
\begin{equation}\label{eq:interactionSS}
 H^{\text{SS}}_{ij} = \frac{16\pi\alpha_s}{9m_i m_j}\tilde{\delta}_\sigma(\bm{r}_{ij}) \bm{s}_i\cdot\bm{s_j},
\end{equation}
is the spin-spin contact hyperfine potential, where $\bm{s}_i$ and $\bm{s}_j$ are the spin of the $i$-th and $j$-th quark, and $\tilde{\delta}_\sigma(\bm{r}_{ij}) = (\frac{\sigma}{\sqrt{\pi}})^3 \text{e}^{-\sigma^2 r^2_{ij}}$ with $\sigma$ being a parameter.
\begin{equation}\label{eq:interactionT}
 H^{\text{T}}_{ij} = \frac{2\alpha_s}{3m_i m_j} \frac{1}{r^3_{ij}}\left( 3\frac{(\bm{s}_i\cdot\bm{r}_{ij})(\bm{s}_j\cdot\bm{r}_{ij})}{r^2_{ij}} - \bm{s}_i\cdot\bm{s}_j \right),
\end{equation}
is the tensor potential. $H^{\text{SO}}_{ij}$ is the spin-orbital interaction potential, 
\begin{eqnarray}\nonumber
 H^{\text{SO}}_{ij}  &=&  \frac{(\bm{s}_i + \bm{s}_j)\cdot\bm{L}_{ij}}{8}\left[ (\frac{1}{m^2_i} + \frac{1}{m^2_j})(\frac{4\alpha_s}{3r^3_{ij}} - \frac{b}{r_{ij}})\right. \\\label{eq:interactionSO+}
 && \left. + \frac{16\alpha_s}{3m_im_jr^3_{ij}} \right],
\end{eqnarray}
where $\bm{L}_{ij} = (\bm{r}_i - \bm{r}_j)\times \frac{m_j \bm{p}_i - m_i \bm{p}_j}{m_i + m_j}$ is the orbital angular momentum between the $i$-th and $j$-th quark, $\bm{p}_{i,j}$ are the momentum of the quarks. The potential containing $\frac{1}{r^3}$ is divergent. A cutoff $R_c$ is introduced, so that $\frac{1}{r^3} \to \frac{1}{R^3_c}$ for $r \leq R_c$. Note that, $V_{ij} = V_{q\bar{q}}/2$, where $V_{q\bar{q}}$ is the potential between the quark and antiquark \cite{Sun2023}.

 To explore the effect of a running coupling constant on the nucleons' elastic form factors, we use the following form \cite{Lakhina2006,Sun2023}
\begin{equation}\label{eq:runningAlphas}
 \alpha_s(Q^2) = \frac{4\pi}{\beta \log(\text{e}^{\frac{4\pi}{\beta\alpha_0}} + \frac{Q^2}{\Lambda^2_{\text{QCD}}})},
\end{equation}
where $\Lambda_{\text{QCD}}$ is the energy scale below which nonperturbative effect dominates, $\beta = 11-\frac{2}{3}N_f$ with $N_f$ being the flavor number, $Q$ is the transferred momentum. Eq. (\ref{eq:runningAlphas}) approaches the one loop running form of perturbative QCD at large $Q^2$ and saturates at low $Q^2$. $\alpha_0$ is the only free parameter in Eq. (\ref{eq:runningAlphas}). 

Derived from the one gluon exchange interaction, the $\alpha_s$'s in equations (\ref{eq:interactionSI})$\sim$(\ref{eq:interactionSO+}) should be equal to one another. While in general they are independent parameters. In this work we treat them equal to one another, except that we tune the interaction of Eq. (\ref{eq:interactionSS}) to see its effect on the mixing state, Eq. (\ref{eq:mixingPN}). 

In the case of nucleons and $\Delta$s, $m_1 = m_2 = m_3 = M_q$, it's convenient to introduce the Jacobi coordinate,
\begin{eqnarray}
 \bm{\rho} &=& \frac{1}{\sqrt{2}} (\bm{r}_1-\bm{r}_2),\\
 \bm{\lambda} &=& \sqrt{\frac{2}{3}} (\frac{\bm{r}_1+\bm{r}_2}{2} - \bm{r}_3).
\end{eqnarray}
So that the kinematic energy in the center of mass frame (CMF) $T =\sum\limits_{i} T_i = -\frac{1}{2M_q} (\nabla^2_{\bm{\rho}} + \nabla^2_{\bm{\lambda}})$, where $\nabla^2_{\bm{\rho}}$ and $\nabla^2_{\bm{\lambda}}$ are the Laplacian operetor on the $\bm{\rho}$ and $\bm{\lambda}$. The orbital angular momentum in CMF is $\bm{L} = M_q(\bm{\rho}\times \bm{\dot{\rho}} + \bm{\lambda}\times \bm{\dot{\lambda}})$.

\subsection{The wavefunction}\label{subsec:wavefunction}

In quark model the wave function of a baryon decomposes,
\begin{equation}
 \Psi = \phi \chi F \psi,
\end{equation}
where $\phi,\, \chi,\, F,\, \psi$ are the color, spin, flavour and space wave functions.
The total wave function $\Psi$ is antisymmetric when any quark pairs interchange, so $\phi,\, \chi,\, F,\, \psi$ should be representations of $S_3$ permutation group. We follow the standard way to represent and classify the wave function \cite{Horgan1973,Giannini1991,Li1991,Liu2020}. The specific form we adopted is presented in Appendix A.

It's mensionable that we use a superposition of $\psi^\sigma_{NLM_L}$, Eq. (\ref{eq:spacialWF}), to represent the spacial wave function, i.e.
\begin{equation}\label{eq:spacialWFsuperPosition}
 \psi^\sigma_{NLM_L}(r,\theta,\varphi) = \sum\limits_{x}C_x \psi^\sigma_{NLM_L}(\alpha_x r,\theta,\varphi).
\end{equation}
$\alpha_x = \alpha_1 d^{x-1}, x = 1,2,\cdots,n$ is a geometric progression with proportion $d$. Physically this means we use a superposition of harmonic oscillator function to mimic the spacial wave function. Mathematically this leads to a mixing between the $|N_6, ^{2S+1}N_3,N,L,J\rangle$ and  $|N_6, ^{2S+1}N_3,N',L,J\rangle$ states. 

The spacial wave function, Eq. (\ref{eq:spacialWFsuperPosition}), has been used to study the $\Omega$ baryon spectrum and decays, which turns out successful \cite{Liu2020}. When studying the nucleon and $\Delta$ spectrum, we assume Eq. (\ref{eq:spacialWFsuperPosition}) which has definite $(N,L)$ value to be the physical state and ignore possible mixing between the states with different $(N_6,L,S)$ but the same $J^P$ value. However, mixings are treated carefully when we study the form factors.

The total wave function could be written as
\begin{equation}
 \Psi = \sum\limits_x C_x\Psi(\alpha_x),
\end{equation}
where $\Psi(\alpha_x) = \phi \chi F \psi(\alpha_x r,\theta,\varphi)$. To solve the eigenvalue and wave function in Eq. (\ref{eq:schrodinger}), we need to calculate the matrix elements of the hamiltonium, $\langle \Psi(\alpha_y)|H |\Psi(\alpha_x) \rangle$. This is most easily done using the Wigner-Eckart theory. The formulae used are listed in Appendix B. Note that because the total wave function is antisymmetric, the matrix elements of the potential satisfy $\langle \Psi(\alpha_y)|V_{13} |\Psi(\alpha_x) \rangle = \langle \Psi(\alpha_y)|V_{23} |\Psi(\alpha_x) \rangle =\langle \Psi(\alpha_y)|V_{12} |\Psi(\alpha_x) \rangle$.

\subsection{The form factors}\label{subsec:formfactor}

Among the states upto $N=2$ (see Table \ref{tab:totalWF}), there are five states with $J^{P} = (\frac{1}{2})^+$, i.e.
\begin{eqnarray}\label{eq:pureSWave}
 |\psi_1 \rangle &=& |56, ^{2}8,0,0,(\frac{1}{2})^{\textmd{+}}\rangle,\\
 |\psi_2 \rangle &=& |56, ^{2}8,2,0,(\frac{1}{2})^{\textmd{+}}\rangle,\\\label{eq:psi3}
 |\psi_3 \rangle &=& |70, ^{2}8,2,0,(\frac{1}{2})^{\textmd{+}}\rangle,\\\label{eq:psi4}
 |\psi_4 \rangle &=& |70, ^{4}8,2,2,(\frac{1}{2})^{\textmd{+}}\rangle,\\
 |\psi_5 \rangle &=& |20, ^{2}8,2,1,(\frac{1}{2})^{\textmd{+}}\rangle.
\end{eqnarray}
In principle all these states could mix and contribute to proton/neutron. There is indeed a reason to consider the mixing. If neutron is composed of the pure S-wave state, $|\psi_1 \rangle$, its electric form factor will be zero, which is in conflict with experiment. The mixing state was proposed by the pioneers long ago \cite{Isgur1978}. 

In the scheme herein $|\psi_2 \rangle$ is the radial excitation of $|\psi_1 \rangle$, mixing between these two states has been considered via Eq. (\ref{eq:spacialWFsuperPosition}). There is no observed state corresponding to $|\psi_5 \rangle$ up to now, so we ignore it. We propose the proton/neutron wave function to be of this form,
\begin{equation}\label{eq:mixingPN}
 |\psi \rangle_{p,n} = c_1|\psi_1 \rangle + c_3|\psi_3 \rangle + c_4|\psi_4 \rangle.
\end{equation}
The parameters $c_1$, $c_3$ and $c_4$ are calculated using the hamiltonian specified in Eqs. (\ref{eq:hamiltonian})$\sim$(\ref{eq:interactionSO+}). Note that the only three nonzero mixing matrix elements are $\langle \psi_3|H^{SS}|\psi_1 \rangle$, $\langle \psi_4|H^{T}|\psi_1 \rangle$, $\langle \psi_4|H^{T}|\psi_3 \rangle$ (and their conjugates). $|\psi_2 \rangle$ is regarded as the roper resonance, ignoring any mixing.

In the nonrelativistic limit, the electric and magnetic form factors describe the distribution of the electric charge and the magnetic momentum. The electric form factor is \cite{Giannini1991},
\begin{equation}\label{eq:electricFF}
G_E(Q^2) =  \langle \psi|\sum\limits_i q_i e^{-i\bm{Q}\cdot\bm{r}_i}|\psi \rangle,
 \end{equation}
where $Q^2 = \bm{Q}^2$, $i = \{u,u,d\}$ for proton and $i = \{u,d,d\}$ for neutron, $q_i$ is the charge number of the quark, $\bm{Q}$ is the momentum transferred. The magnetic form factor is \cite{Giannini1991},
\begin{equation}\label{eq:magneticFF}
G_M(Q^2) = \langle \psi|\sum\limits_i \mu_{iz} e^{-i\bm{Q}\cdot\bm{r}_i}|\psi \rangle,
 \end{equation}
 where $\mu_{iz} = \frac{q_i e\hbar}{2m_i}\sigma_{iz}$ is the magnetic momentum of the constituent quark. In the isospin symmetry limit, $m_u = m_d = M_q$. The constituent quark mass
$M_q$ is usually considered as a constant. However the quarks inside the nucleon are not real particles, and they are off the mass shell generally. In the study of Dyson-Schwinger equations and lattice QCD, an off mass shell quark possesses a running quark mass $M_q=M_q(p^2)$ \cite{Bhagwat2007a}, where $p$ is the off shell four momentum of the quark. We assume
\begin{equation}\label{eq:Q2p2Ratio}
 Q^2 = \xi p^2,
\end{equation}
where $\xi$ is a parameter. The $Q^2$ dependence of $|\psi \rangle$, which means boosting the wave function , has a weak effect \cite{Sun2023}, so it is ignored herein. Then the $Q^2$ dependence of the magnetic form factor factorizes
\addtocounter{equation}{-2}
\renewcommand{\theequation}{\arabic{equation}'}
\begin{equation}\label{eq:magneticFFMQ}
 G_M(Q^2) = \frac{ e\hbar}{2M_q(\frac
{Q^2}{\xi})} \cdot \langle \psi|\sum\limits_i q_i \sigma_{iz} e^{-i\bm{Q}\cdot\bm{r}_i}|\psi \rangle.
\end{equation}
\addtocounter{equation}{1}
\renewcommand{\theequation}{\arabic{equation}}
The radius is determined by
\begin{equation}
 r_{E,M} = -6\frac{G'_{E,M}(0)}{G_{E,M}(0)}.
\end{equation}

\section{Results}\label{sec:results}

\subsection{Nucleon and $\Delta$ spectrum}\label{subsec:spectrum}

\begin{table}[h!]\centering
\caption{\label{tab:parameters} Two parameter sets used in our calculation.}
\begin{tabular}{c|c|c|c|c|c|c}
\hline    
 & $M_q/\text{GeV}$ & $\alpha_S$ or $\alpha_0$ & $b/\text{GeV}$ & $\sigma/\text{GeV}^2$ & $R_c/\text{fm}$ & $C_0/\text{GeV}$\\
\hline    
constant $\alpha_S$ & 0.21 & 0.863 & 0.10 & 0.94 & 0.37 & -0.461\\
\hline
running $\alpha_S$ & 0.21 &  0.900 & 0.11 & 1.20 & 0.50 & -0.687\\
\hline
\end{tabular}
\end{table}

\begin{widetext}

\begin{table}[ht!]\centering
\caption{\label{tab:nucleon} Masses (in MeV) of nucleons under 2000 MeV. The first column is the state classfied in quark model, with $J^P = (\frac{1}{2})^+$ states marked by an arrow. The second and third columns are our results calculated with the two groups of parameters in Table \ref{tab:parameters}. The fourth column is the result of Capstick and Isgur \cite{Capstick1986}, which takes into account the relativistic correction. The fifth column is the name and status of the probable corresponding state in particle data group (PDG), and the last column is the observed value \cite{Workman2022}. For three-star and four-star states, the range of the observed value is cited; for one-star and two-star states, we just cite an approximate value.}
\begin{tabular}{l|c|c|c|l|c}
\hline    
$|N_6, ^{2S+1}N_3,N,L,J^{\textmd{P}}\rangle$  & constant $\alpha_s$ & running $\alpha_s$ & CI \cite{Capstick1986} & PDG name &expt. \cite{Workman2022} \\
\hline
$|56, ^{2}8,0,0,(\frac{1}{2})^{\textmd{+}}\rangle \leftarrow$ & 963 & 951 & 960 & p,n $^{****}$ & 939\\
$|70, ^{2}8,1,1,(\frac{1}{2})^{\textmd{-}}\rangle$ & 1485 & 1429 & 1460 & N(1535)$^{****}$ & 1500-1520 \\
$|70, ^{2}8,1,1,(\frac{3}{2})^{\textmd{-}}\rangle$ & 1531 & 1425 & 1495 & N(1520)$^{****}$ & 1505-1515\\
$|70, ^{4}8,1,1,(\frac{1}{2})^{\textmd{-}}\rangle$ & 1524 & 1471 & 1535 & N(1650)$^{****}$ & 1640-1670 \\
$|70, ^{4}8,1,1,(\frac{3}{2})^{\textmd{-}}\rangle$ & 1660 & 1543 & 1625 & N(1700)$^{***}$& 1650-1750 \\
$|70, ^{4}8,1,1,(\frac{5}{2})^{\textmd{-}}\rangle$ & 1620 & 1473 & 1630 & N(1675)$^{****}$ & 1655-1665 \\
$|56, ^{2}8,2,0,(\frac{1}{2})^{\textmd{+}}\rangle\leftarrow$ & 1583 & 1580 & 1540 & N(1440)$^{****}$& 1360-1380\\
$|70, ^{2}8,2,0,(\frac{1}{2})^{\textmd{+}}\rangle\leftarrow$ & 1688 & 1736 & 1770 & N(1710)$^{****}$ & 1680-1720 \\
$|70, ^{4}8,2,0,(\frac{3}{2})^{\textmd{+}}\rangle$ & 1921 & 1809 & 1795 & N(1900)$^{****}$& 1900-1940\\
$|20, ^{2}8,2,1,(\frac{1}{2})^{\textmd{+}}\rangle\leftarrow$ & 1948& 1861 & 1880 &  &  \\
$|20, ^{2}8,2,1,(\frac{3}{2})^{\textmd{+}}\rangle$ & 1952 & 1841 & 1870&  &  \\
$|56, ^{2}8,2,2,(\frac{3}{2})^{\textmd{+}}\rangle$ & 1877 & 1803 & 1910& N(1720)$^{****}$ & 1660-1690 \\
$|56, ^{2}8,2,2,(\frac{5}{2})^{\textmd{+}}\rangle$ &  1820 & 1725 & 1770 & N(1680)$^{****}$& 1665-1680 \\
$|70, ^{2}8,2,2,(\frac{3}{2})^{\textmd{+}}\rangle$ & 1910 & 1835 & 1950 &  &\\
$|70, ^{2}8,2,2,(\frac{5}{2})^{\textmd{+}}\rangle$ & 1886& 1781 & 1980 & N(1860)$^{**}$ & $\approx 1860$\\
$|70, ^{4}8,2,2,(\frac{1}{2})^{\textmd{+}}\rangle\leftarrow$ & 1910  & 1883 & 1975 & N(1880)$^{***}$ & 1820-1900 \\
$|70, ^{4}8,2,2,(\frac{3}{2})^{\textmd{+}}\rangle$ & 1975 & 1900 & 2030 & N(2040)$^{*}$ & $\approx 2040$\\
$|70, ^{4}8,2,2,(\frac{5}{2})^{\textmd{+}}\rangle$ & 1976 & 1871 & 1995& N(2000)$^{**}$ & $\approx 2000$\\
$|70, ^{4}8,2,2,(\frac{7}{2})^{\textmd{+}}\rangle$ & 1883& 1751 & 1910 & N(1900)$^{**}$& $\approx 1900$\\
$|56, ^{2}8,3,1,(\frac{1}{2})^{\textmd{-}}\rangle$ & 2096 & 2044 & 1945 & N(1895)$^{***}$ & 1890-1930 \\
$|56, ^{2}8,3,1,(\frac{3}{2})^{\textmd{-}}\rangle$ & 2058 & 1996 & 1960 & N(1875)$^{***}$ & 1850-1950 \\
\hline
\end{tabular}
\end{table}

\vspace{-2em}
\begin{table}[ht!]\centering
\caption{\label{tab:delta} Masses (in MeV) of $\Delta$ baryons under 2000 MeV. The meaning of the first row is the same as Table \ref{tab:nucleon}.}
\begin{tabular}{l|c|c|c|l|c}
\hline
$|N_6, ^{2S+1}N_3,N,L,J^{\textmd{P}}\rangle$  & constant $\alpha_s$ & running $\alpha_s$ & CI \cite{Capstick1986}& PDG name &expt. \cite{Workman2022} \\
\hline
$|56, ^{4}10,0,0,(\frac{3}{2})^{\textmd{+}}\rangle$ & 1298 & 1145 & 1230 & $\Delta(1232)^{****}$& 1209-1211 \\
$|70, ^{2}10,1,1,(\frac{1}{2})^{\textmd{-}}\rangle$ & 1623& 1497 & 1555 & $\Delta(1620)^{****}$ & 1590-1610\\
$|70, ^{2}10,1,1,(\frac{3}{2})^{\textmd{-}}\rangle$ & 1623& 1497 & 1620 & $\Delta(1700)^{****}$ & 1640-1690\\
$|56, ^{4}10,2,0,(\frac{3}{2})^{\textmd{+}}\rangle$ & 1863 & 1746 & 1795 & $\Delta(1600)^{****}$ & 1460-1560\\
$|70, ^{2}10,2,0,(\frac{1}{2})^{\textmd{+}}\rangle$ & 1916 & 1806 & 1835 & $\Delta(1750)^{*}$ & $\approx 1750$\\
$|56, ^{4}10,2,2,(\frac{1}{2})^{\textmd{+}}\rangle$ & 2023 & 1938 & 1875 & $\Delta(1910)^{****}$ & 1830-1890\\
$|56, ^{4}10,2,2,(\frac{3}{2})^{\textmd{+}}\rangle$ & 2009 & 1914 & 1915& $\Delta(1920)^{***}$ & 1850-1950\\
$|56, ^{4}10,2,2,(\frac{5}{2})^{\textmd{+}}\rangle$ & 1954 & 1846 & 1910 & $\Delta(2000)^{**}$ & $\approx 2000$\\
$|56, ^{4}10,2,2,(\frac{7}{2})^{\textmd{+}}\rangle$ & 1839 & 1713 &1940 & $\Delta(1950)^{****}$ & 1870-1890\\
$|70, ^{2}10,2,2,(\frac{3}{2})^{\textmd{+}}\rangle$ & 1977 & 1879 &1985 & &\\
$|70, ^{2}10,2,2,(\frac{5}{2})^{\textmd{+}}\rangle$ & 1911 & 1800 & 1990& $\Delta(1905)^{****}$ & 1770-1830\\
$|56, ^{4}10,3,1,(\frac{1}{2})^{\textmd{-}}\rangle$ & 2270 & 2183 & 2035 & $\Delta(1900)^{***}$ & 1830-1900\\
$|56, ^{4}10,3,1,(\frac{3}{2})^{\textmd{-}}\rangle$ & 2220 & 2132 &2080 & $\Delta(1940)^{**}$ & $\approx 1950$\\
$|56, ^{4}10,3,1,(\frac{5}{2})^{\textmd{-}}\rangle$ & 2150 & 2046 & 2155 & $\Delta(1930)^{***}$ & 1840-1920\\
\hline
\end{tabular}
\end{table}
\end{widetext}

We use two parameter sets to calculate our results. One assumes the $\alpha_s$ in equation (\ref{eq:interactionSI})$\sim$(\ref{eq:interactionSO+}) is a constant. The other assumes $\alpha_s$ takes the form, Eq. (\ref{eq:runningAlphas}). We keep $\Lambda_{\text{QCD}} = 300\text{ MeV}$ and $N_f = 2$ in our calculation. There are six parameters in each case: $M_q$, $\alpha_S \text{ or } \alpha_0$, $b$, $\sigma$, $R_c$ and $C_0$, which are listed in Table \ref{tab:parameters}. The parameters are tuned to produce a reasonable nucleon and $\Delta$ spectrum. Especially we try to keep the masses of $\psi_1$, $\psi_3$, $\psi_4$ (defined in Eqs. (\ref{eq:pureSWave},\ref{eq:psi3},\ref{eq:psi4})) consistent with experiment.

The results of the nucleon and $\Delta$ spectrum are listed in Table \ref{tab:nucleon} and Table \ref{tab:delta}. The second and third column of these two tables are the masses of the pure state, the mixing effect may changes the masses $10 \sim 20$ MeV. The fifth column is the name and status of the most probable observed counterpart. The sixth column is its observed value. For three-star and four-star states, the range of the observed value is cited; for one-star and two-star states, we just cite an approximate value \cite{Workman2022}. We also list the result of Capstick and Isgur \cite{Capstick1986} in the fourth column, which takes into account relativistic correction.

The nucleon spectrum almost keeps the band structure of the harmonic oscillator. The $(N,L)=(0,0)$ state is the main constituent of the proton/neutron. The five $(N,L)=(1,1)$ states contain a doublet and a triplet, their masses are in the range $1400\sim 1700 \text{ MeV}$. $|\psi_2\rangle =|56, ^{2}8,2,0,(\frac{1}{2})^{\textmd{+}}\rangle$ is the lightest in the $N=2$ band. It is the most natural counterpart of the roper resonance. The roper resonance is the lightest excited nucleon experimentally, while it is heavier than some of the $(N,L)=(1,1)$ states theoretically. This is known as the level ordering problem. We are not going to solve this problem in this paper. We assume $|\psi_2\rangle$ is the roper resonance and predict its elastic form factors and its charge radius in subsection \ref{subsec:GERGMR}. Upto 2000 MeV the nucleons in quark model and the observed ones have a one-to-one correspondence except that one $J^P = (\frac{1}{2})^+$ state and two $J^P = (\frac{3}{2})^+$ states are missing. A similar discussion on the $\Delta$ spectrum could be made. 

The running coupling effect on the mass spectrum is obvious by comparing the second and third columns in Table \ref{tab:nucleon} and \ref{tab:delta}. The deviation is less than 100 MeV for most of them. The largest deviation is about 150 MeV ($|70, ^{4}8,1,1,(\frac{5}{2})^{\textmd{-}}\rangle$ and $|56, ^{4}10,0,0,(\frac{3}{2})^{\textmd{+}}\rangle$). Comparing the theoretical and the experimental results, the errors are less than 150 MeV for most of them. The largest error is about 200 MeV ($|56, ^{2}8,2,0,(\frac{1}{2})^{\textmd{+}}\rangle$ and $|56, ^{4}10,2,0,(\frac{3}{2})^{\textmd{+}}\rangle$). Considering that all these 35 masses are calculated using only 6 free parameters, the spectrum is not bad.

\subsection{Electric form factors of proton and neutron}\label{subsec:GEpGEn}

\begin{table}[ht!]\centering
\caption{\label{tab:mixing} The mass (in MeV) and mixing coefficients of the ground nucleon, assuming Eq. (\ref{eq:mixingPN}).}
\begin{tabular}{c|c|c|c|c}
\hline    
 & $m_{p,n}$& $c_1$ & $c_3$ & $c_4$\\
\hline    
constant $\alpha_S$ & 939 & 0.986 & 0.149 & 0.0679\\
\hline
running $\alpha_S$ & 939 & 0.993 & 0.109 & 0.0430\\
\hline
\end{tabular}
\end{table}

\begin{figure}[ht!]
\centering
 \includegraphics[width=0.485\textwidth]{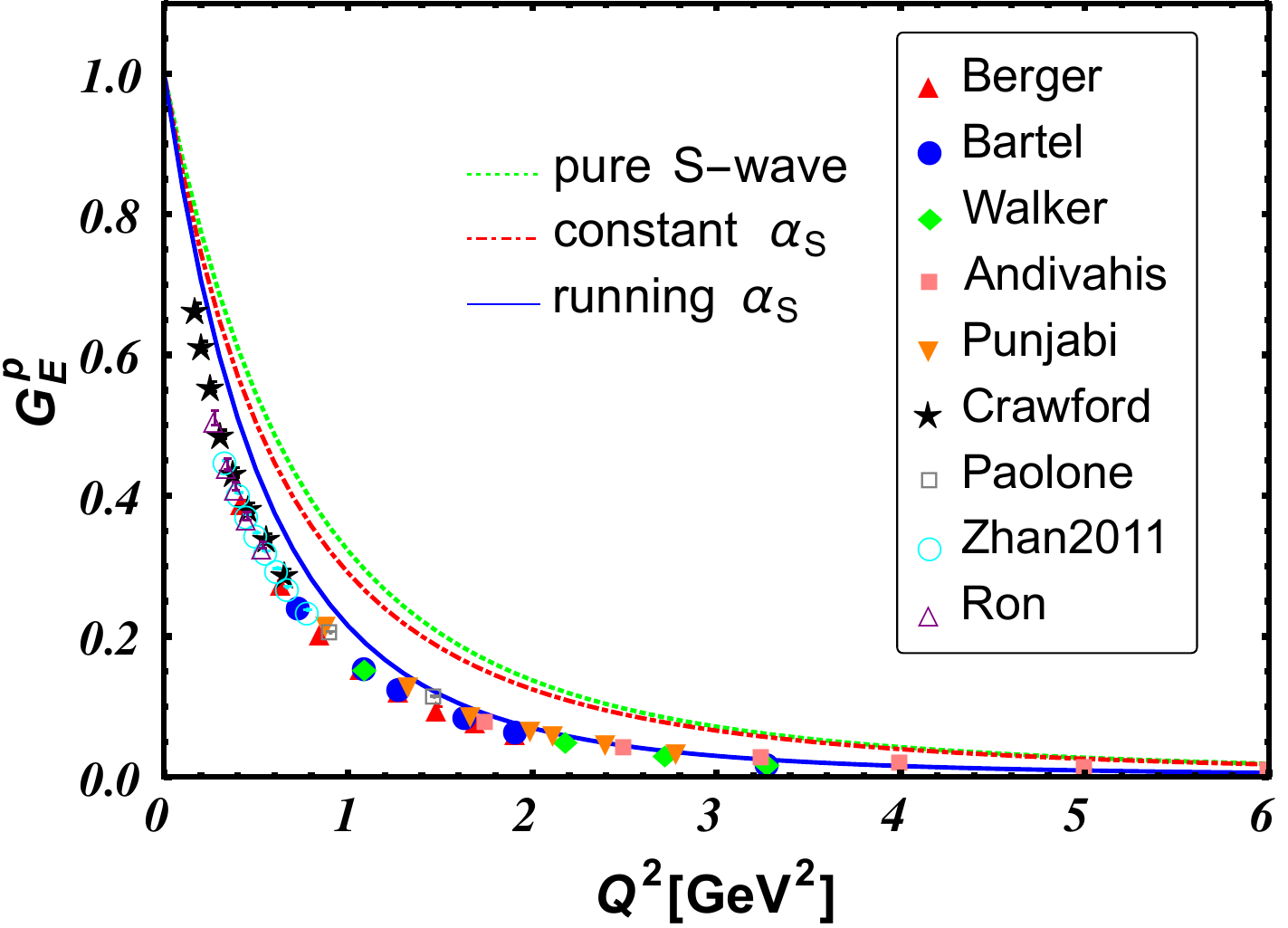}
 \includegraphics[width=0.485\textwidth]{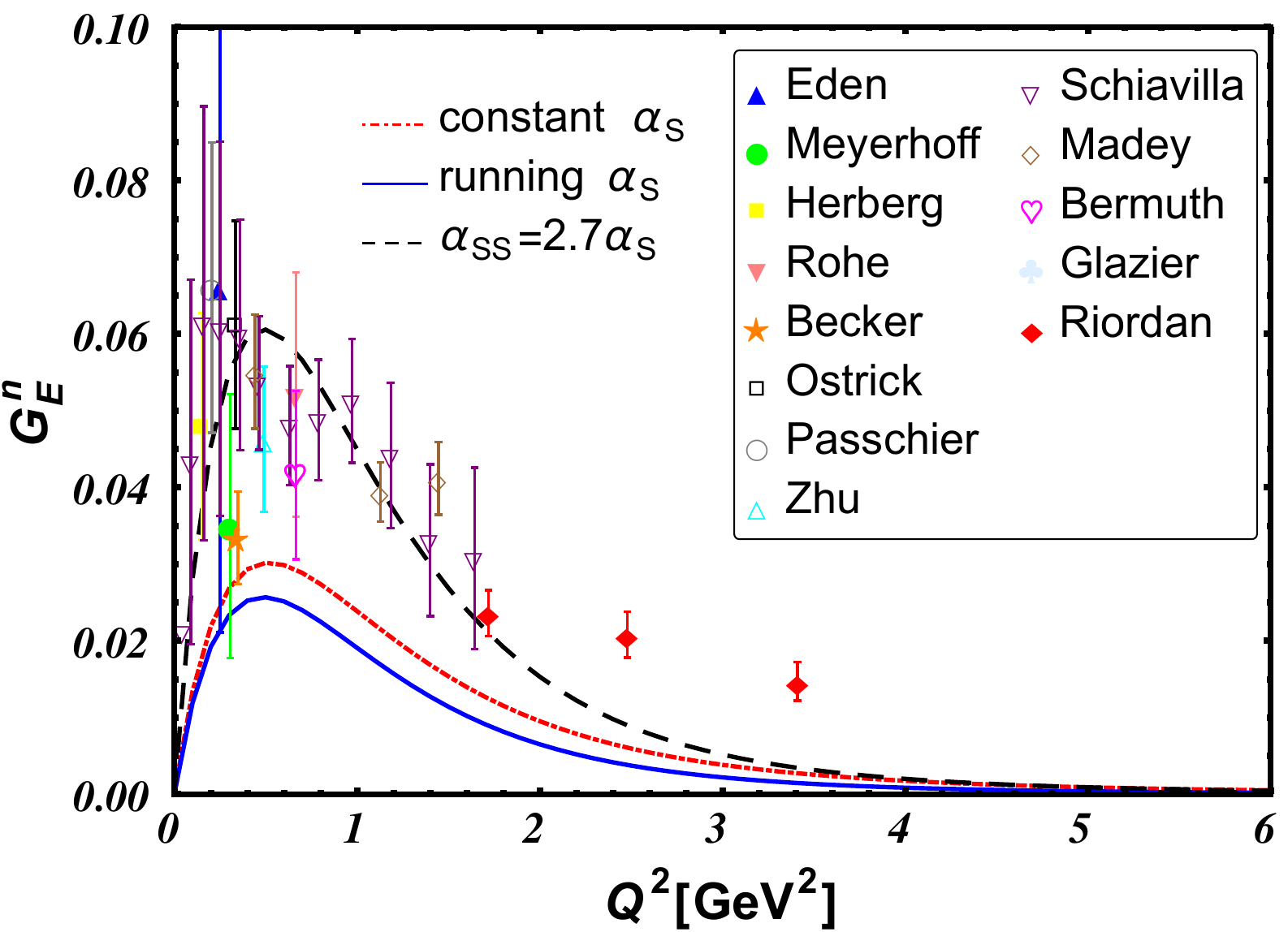}
 \caption{\label{fig:GEpn} (Colored Online) The electric form factor of the proton (upper) and neutron (lower). The red dot-dashed lines are the results calculated using the parameter set with a constant $\alpha_S$. The blue solid lines are the results calculated using the parameter set with a running $\alpha_S$. The green dotted line in the upper panel is the proton electric form factor using the pure S-wave component, Eq. (\ref{eq:pureSWave}), with a constant $\alpha_S$. The black dashed line in the lower panel is the neutron electric form factor after tuning the coupling constant of the scalar hyperfine interaction, Eq. (\ref{eq:interactionSS}), to be $\alpha_{SS}=2.7\alpha_{S}$, with a running $\alpha_S$. The experiment electric form factor values of the proton are from Ref. \cite{Berger1971,Bartel1973,Andivahis1994,Walker1994,Punjabi2005,Crawford2007,Paolone2010,Zhan2011}, those of the neutron are from Ref. \cite{Lung1993,Eden1994,Meyerhoff1994,Becker1999,Herberg1999,Ostrick1999,Passchier1999,Rohe1999,Schiavilla2001,Zhu2001,Bermuth2003,Madey2003,Glazier2005,Riordan2010}, all of which are labelled by the family name of the first author of the paper.}
\end{figure}

The mass of the mixing state, Eq. (\ref{eq:mixingPN}), is 939 MeV, which equals the observed value. The mixing coefficients correspond to the two parameter sets in Table \ref{tab:parameters} are listed in Table \ref{tab:mixing}. The mixing is weak, however it's crucial to produce a nonzero neutron electric form factor. In fact, only the $|\psi_1 \rangle$ and $|\psi_3 \rangle$ cross term and $|\psi_4 \rangle$ itself contribute to the neutron electric form factor,
\begin{eqnarray}\nonumber
G^n_E(Q^2) &= &2c_1c_3 \langle \psi_1|\sum\limits_i q_i e^{-i\bm{Q}\cdot\bm{r}_i}|\psi_3 \rangle \\\label{eq:electricFFn}
&&+ c_4^2 \langle \psi_4|\sum\limits_i q_i e^{-i\bm{Q}\cdot\bm{r}_i}|\psi_4 \rangle.
 \end{eqnarray}
As $c_4$ is very small, $G^n_E(Q^2)$ is mainly determined by the cross term, or by the scalar hyperfine interaction, Eq. (\ref{eq:interactionSS}).

The neutron electric form factor is shown in the lower panel of Figure \ref{fig:GEpn}. The red dot-dashed line is calculated using the parameter set with a constant $\alpha_S$, the blue solid line is calculated using the parameter set with a running $\alpha_S$. They both have a bump near $Q^2 = 0.5 \text{ GeV}^2$. It's consistent with the observed value qualitatively except that the bump is too short. If we treat the coupling strength in Eq. (\ref{eq:interactionSS}) as a free parameter, increasing it to $\alpha_{SS} = 2.7 \alpha$, then the bump (the dashed black line) matches the observed value. However it seems that $G^n_E(Q^2)$ decreases too fast as $Q^2 > 2\text{ GeV}^2$.
 
 The proton electric form factor results are in the upper panel of Figure \ref{fig:GEpn}. The green dotted line is the result of the pure S-wave state, Eq. (\ref{eq:pureSWave}), using a constant $\alpha_S$. The red dot-dashed line is the result of the mixing state using a constant $\alpha_S$, the blue solid line is the result of the mixing state using a running $\alpha_S$. The mixing has a 
weak effect on $G^p_E(Q^2)$, while a running $\alpha_S$ softens $G^p_E(Q^2)$ signficantly and the result matches the observed value much better.

\subsection{Magnetic form factors of proton and neutron}\label{subsec:GMpGMn}

\begin{figure}[t!]
\centering
 \includegraphics[width=0.485\textwidth]{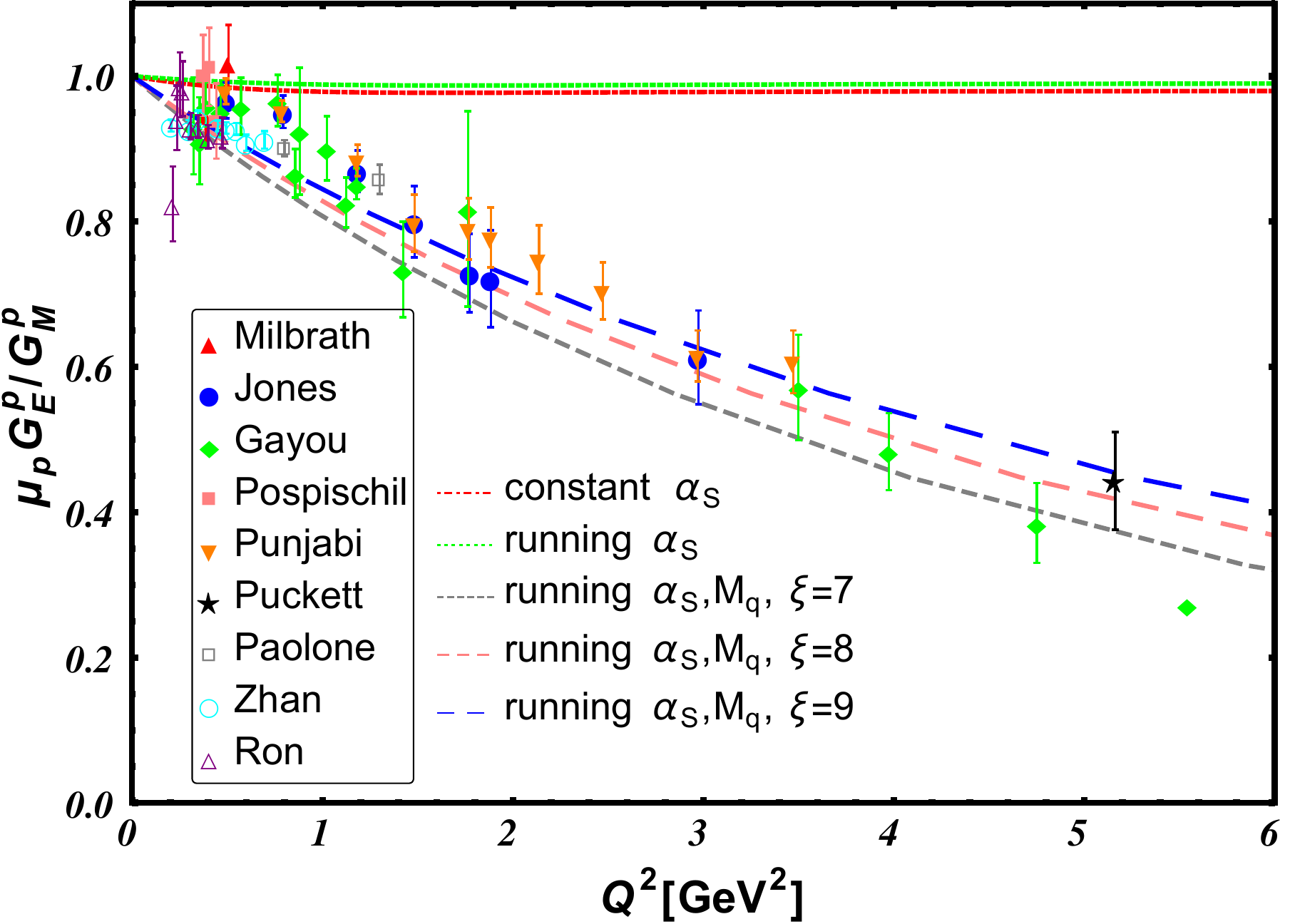}
 \caption{\label{fig:GEpTGMp} (Colored Online) The ratio of the electric and magnetic form factor of the proton, $\mu_p G_E^p/G_M^p$. The green dotted line is calculated using the parameter set with a constant $\alpha_S$ in Table \ref{tab:parameters}, and the red dot-dashed line is calculated using the parameter set with a running $\alpha_S$, both ignoring the running quark mass effect. The three dashed lines are calculated using the parameter set with a running $\alpha_S$, considering the running quark mass effect. The short-dashed gray, middle-dashed pink, long-dashed blue lines correspond $\xi = 7,8,9$ in Eq. (\ref{eq:Q2p2Ratio}). The experiment values are from Ref. \cite{Jones2000,Gayou2001,Pospischil2001,Gayou2002,Punjabi2005,Paolone2010,Puckett2010,Ron2011,Zhan2011}, labelled by the family name of the first author of the paper.}
\end{figure}

\begin{figure}[t!]
\centering
 \includegraphics[width=0.485\textwidth]{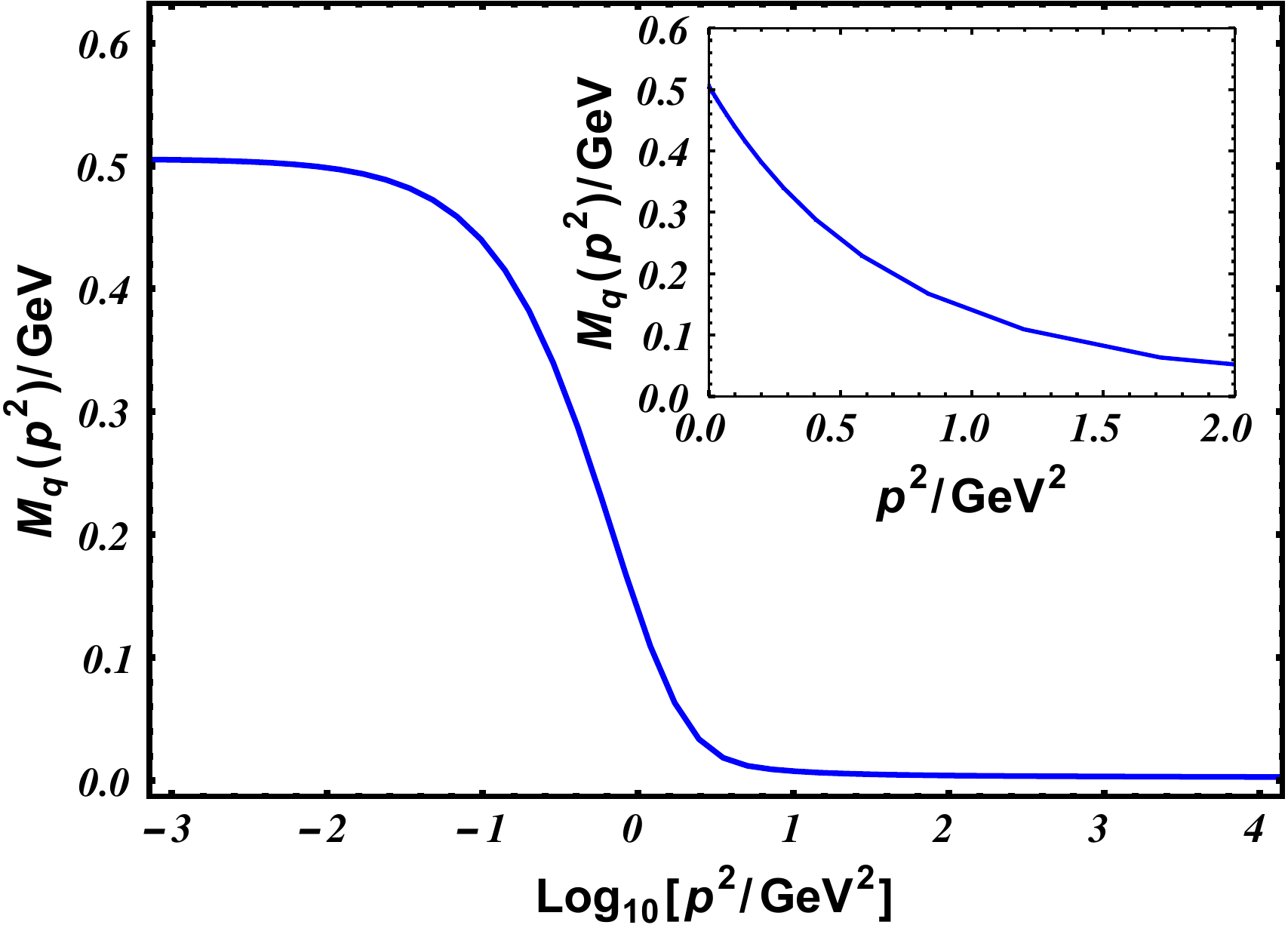}
 \caption{\label{fig:Mq} (Colored Online) The running quark mass function of the $u/d$ quark from solving the Dyson-Schwinger equation of the quark propagator. To solve the equation, an realistic interaction model was used \cite{Chen2018,Chen2019}. The inlaid figure shows the detail in the chiral phase transition region.}
\end{figure}

\begin{figure}[ht!]
\centering
 \includegraphics[width=0.485\textwidth]{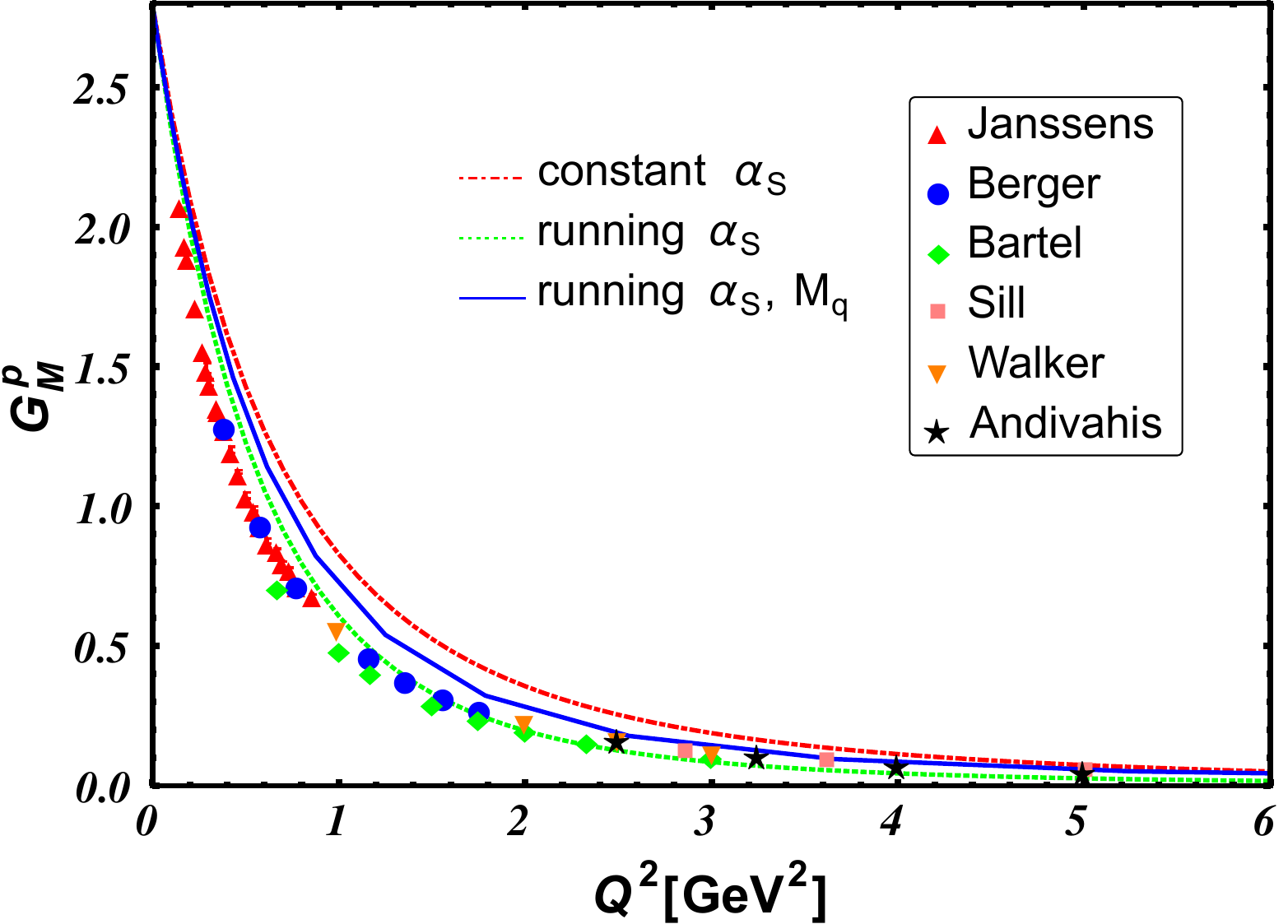}
 \includegraphics[width=0.485\textwidth]{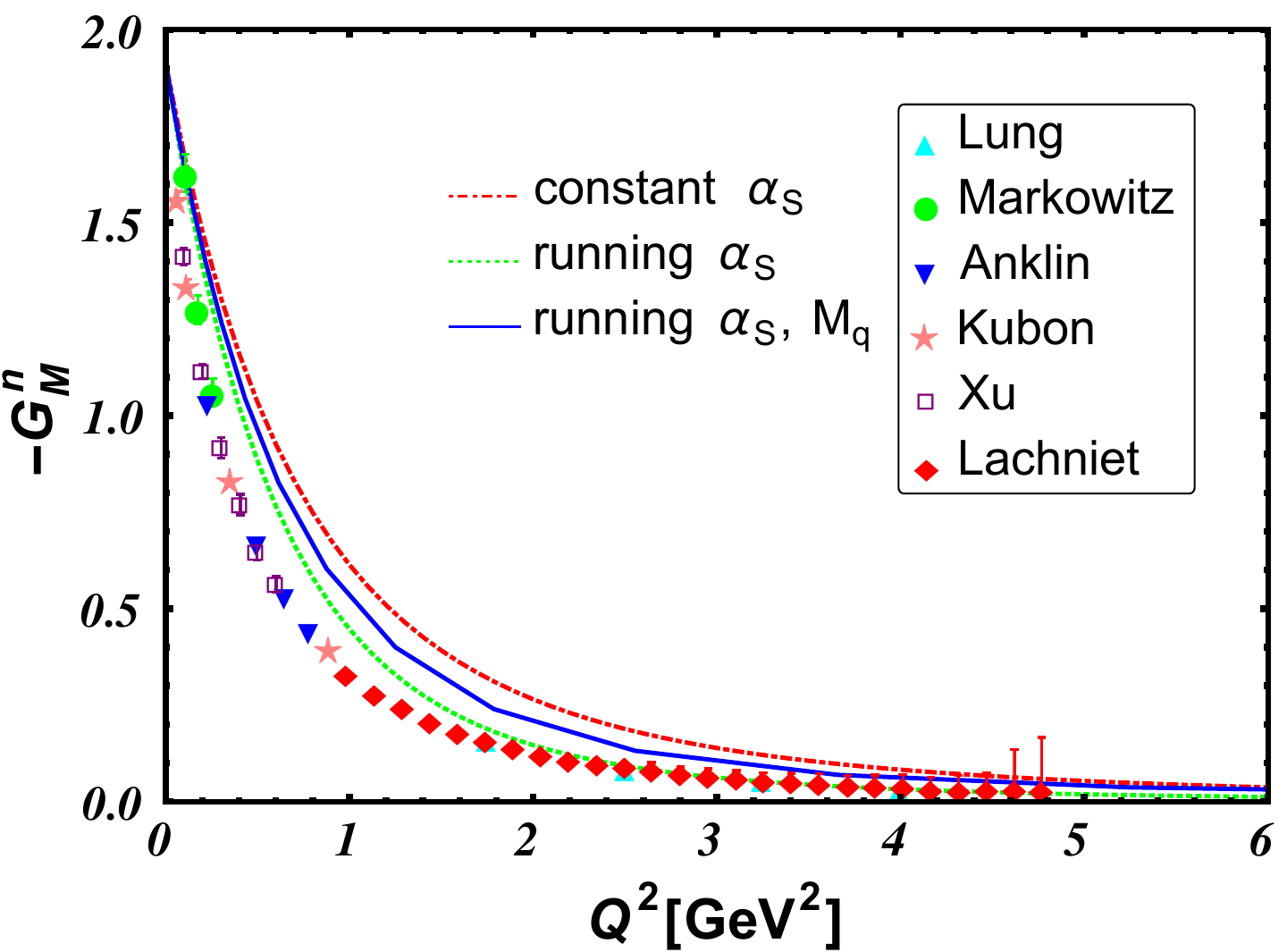}
 \caption{\label{fig:GMpn} (Colored Online) The magnetic form factor of the proton (upper) and neutron (lower). The red dot-dashed lines are the results calculated using the parameter set with a constant $\alpha_S$. The green dotted lines are the results calculated using the parameter set with a running $\alpha_S$. Neither the red dot-dashed line nor the green dotted line considers the running quark mass effect. The blue solid lines are the results calculated using the parameter set with a running $\alpha_S$, taking into account the running quark mass effect. The experiment magnetic form factor values of the proton are from Refs. \cite{Janssens1966,Berger1971,Bartel1973,Sill1993,Andivahis1994,Walker1994}, those of the neutron are from Refs. \cite{Rock1982,Lung1993,Markowitz1993,Anklin1994,Xu2000,Kubon2002,Xu2003,Lachniet2009}, all of which are labelled by the family name of the first author of the paper.}
\end{figure}

In traditional quark model, the constituent quark mass is a constant. The pure S-wave state, Eq. (\ref{eq:magneticFF}), leads to 
\begin{equation}
 G^p_M(Q^2) = \frac{e\hbar}{2M_q} \langle \psi_1|\sum\limits_i q_i e^{-i\bm{Q}\cdot\bm{r}_i}|\psi_1 \rangle.
\end{equation}
$G^p_M(0) = \frac{e\hbar}{2M_q} = \mu_p$ is the magnetic momentum of the proton. $M_q$ is not well constrained, so we can not predict $ \mu_p$ precisely. Herein we normalize $G^p_M(0)$ using the experimental value of $\mu_p$. A  pure S-wave state leads to $ \mu_p G_E^p/G_M^p =1$. As the mixing is weak, we still have $ \mu_p G_E^p/G_M^p \approx1$ with a mixing state. The result of $ \mu_p G_E^p/G_M^p$ calculated using a constant $\alpha_S$ is the red dot-dashed line in Figure \ref{fig:GEpTGMp}, the result calculated using a running $\alpha_S$ is the green dotted line. They both are equal to one approximately, which is in contradiction with the more undisputable polarization measurement.

If we consider a running quark mass, the magnetic form factor factorizes as Eq. (\ref{eq:magneticFFMQ}). Then $\mu_p G_E^p/G_M^p \approx \frac{M_q(p^2=Q^2/\xi)}{M_q(0)}$. $M_q(p^2)$ is the running quark mass function. We obtain $M_q(p^2)$ by solving the Dyson-Schwinger equation of the quark propagator, which is shown in Figure \ref{fig:Mq}. To solve the equation, an realistic interaction model was used \cite{Chen2018,Chen2019}. $M_q(p^2)$ approaches the on shell constituent quark mass when $p^2 \to 0$, and approaches the current quark mass when $p^2 \to \infty$. $M_q(p^2)$ undergoes a phase transition near $p^2 \approx 1 \text{ GeV}^2$.

We could have a rough estimation of $\xi$. When $|\bm{Q}|$ is small, it is equally distributed on the three quarks, so $p^2 = Q^2/9$. When $|\bm{Q}|$ increases, it is mainly transferred to the struck quark, we expect $p^2 > Q^2/9$. Using Eq. (\ref{eq:magneticFFMQ}) for $G_M^p$, $\mu_p G_E^p/G_M^p$ are calculated with $\xi = 7,8,9$, which are the short, middle and long dashed lines in Figure \ref{fig:GEpTGMp}. $\frac{M_q(Q^2/9)}{M_q(0)}$ matches the observed value of $\mu_p G_E^p/G_M^p$ upto $Q^2 \approx 4\text{ GeV}^2$ perfectly. When $Q^2 > 4\text{ GeV}^2$, observed $\mu_p G_E^p/G_M^p$ approaches $\frac{M_q(Q^2/\xi)}{M_q(0)}$ with a smaller $\xi$ as expected.

It is noteworthy that relativistic correction is not beneficial to reconcile the theoretical and observed value of $\mu_p G_E^p/G_M^p$ \cite{DeSanctis2000,DeSanctis2007}. Physically a running quark mass means the quark is not a point particle. In fact, $\mu_p G_E^p/G_M^p$ could be phenomenologically well fitted by introducing a quark form factor \cite{DeSanctis2007,Gross2008}. Herein we show that the quark mass function, $M_q(Q^2/\xi)$, is responsible for the decreasing of $\mu_p G_E^p/G_M^p$ as $Q^2$ increases.

Our results for the magnetic form factor of proton and neutron are shown in Figure \ref{fig:GMpn}. The neutron magnetic form factor is normalized using the experiment value $G_M^n(0) = \mu_n$. The red dot-dashed lines are the results calculated using the parameter set with a constant $\alpha_S$. The green dotted lines are the results calculated using the parameter set with a running $\alpha_S$. Neither the red dot-dashed line nor the green dotted line considers the running quark mass effect. The blue solid lines are the results calculated using the parameter set with a running $\alpha_S$, taking into account the running quark mass effect. The blue solid lines are our final predictions for $G_M^p$ and $G_M^n$. Considering $G_E^p$, $G_M^p$ and $G_M^n$ together, our results are satisfactory except that they are a little larger in the phase transition region: $G_E^p$, $0.5\text{ GeV}^2\lesssim Q^2 \lesssim 1.2\text{ GeV}^2$; $G_M^p$, $0.4\text{ GeV}^2\lesssim Q^2 \lesssim 2.5\text{ GeV}^2$; $G_M^n$, $0.3\text{ GeV}^2\lesssim Q^2 \lesssim 3.5\text{ GeV}^2$. This may indicate the lack of goldstone boson exchange interaction \cite{Wagenbrunn2001} and it is to be studied further.

\subsection{Elastic form factors of the charged roper resonance}\label{subsec:GERGMR}

\begin{figure}[t!]
\centering
 \includegraphics[width=0.485\textwidth]{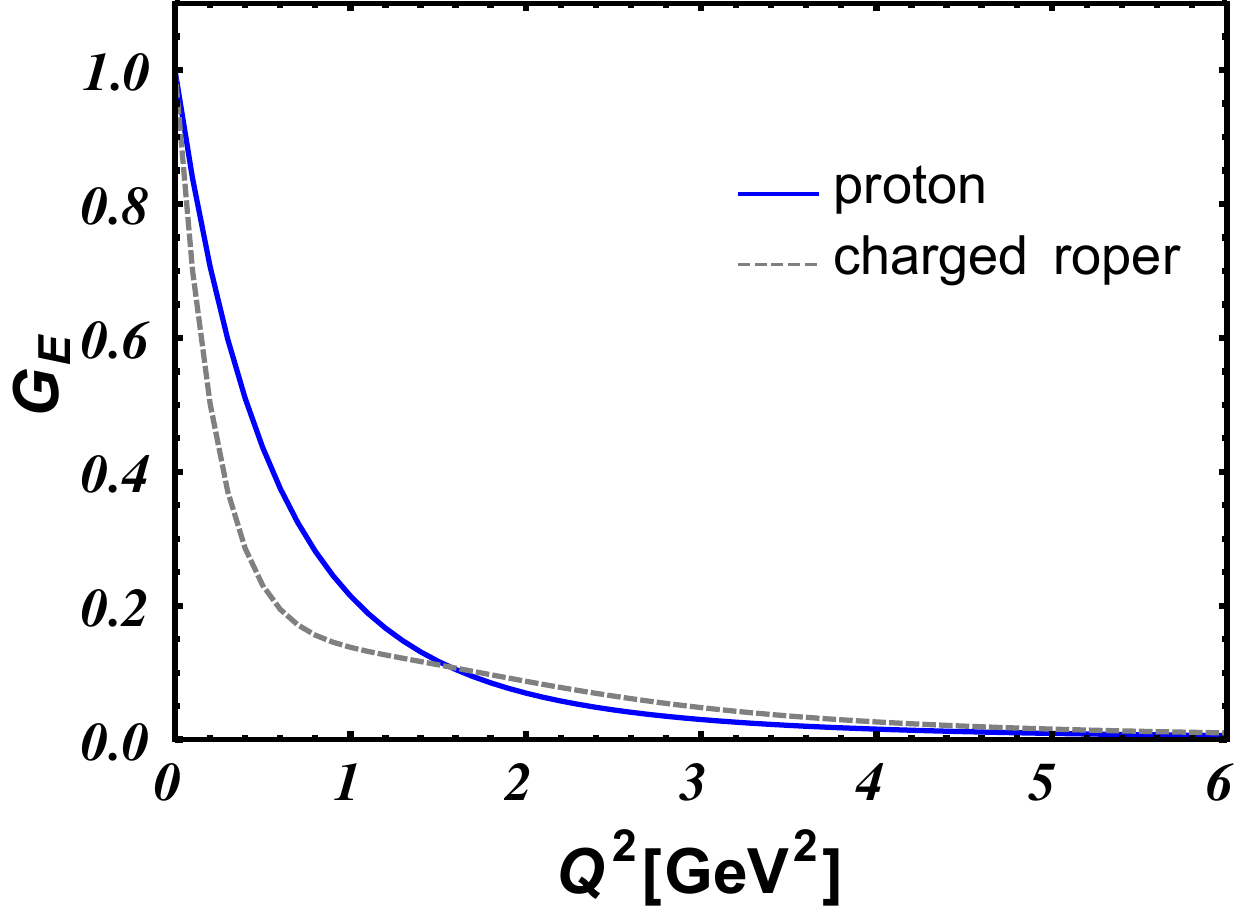}
 \includegraphics[width=0.485\textwidth]{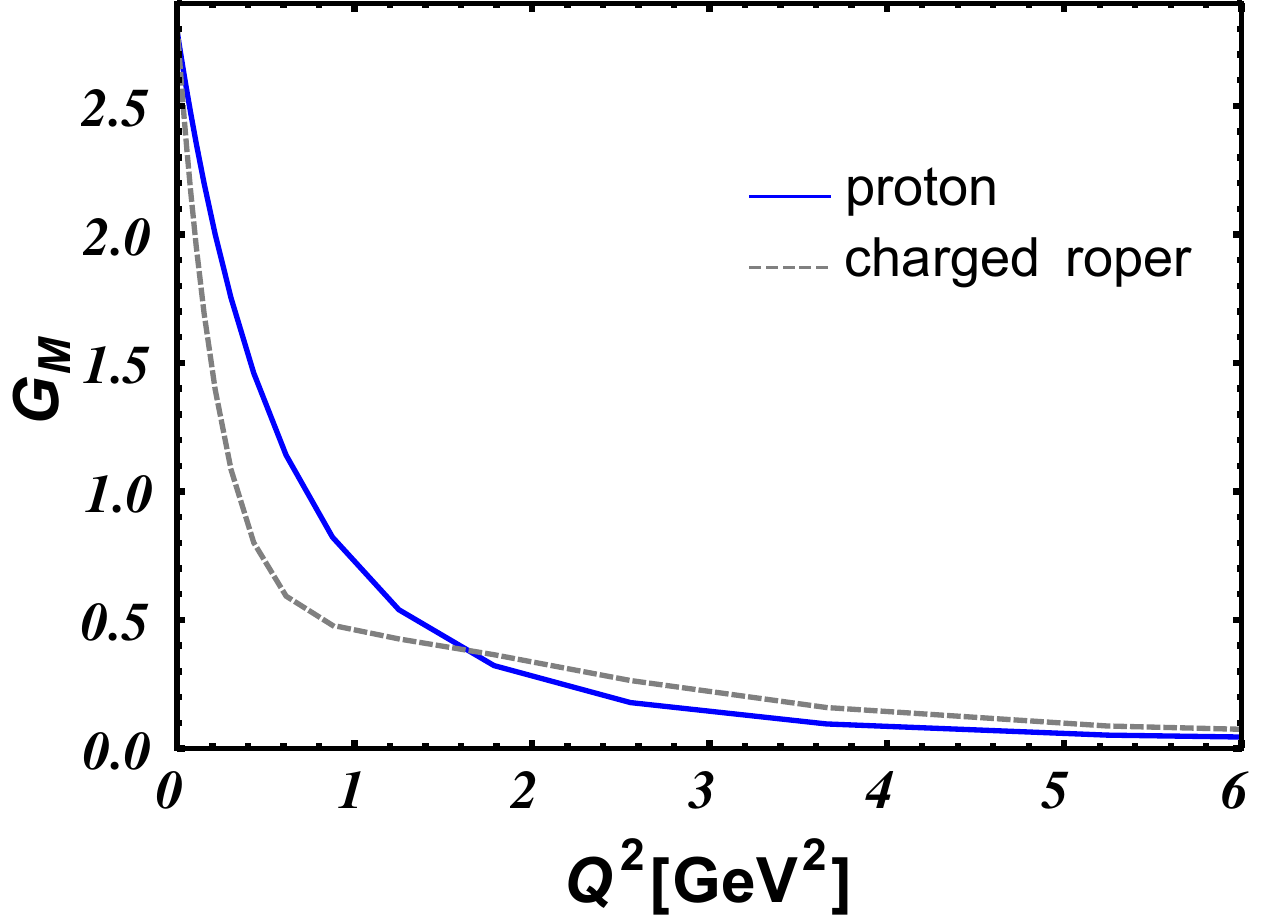}
 \caption{\label{fig:GEMR} (Colored Online) The electric (upper) and magnetic (lower) form factor of the charged roper resonance. Those of the proton are displayed for comparison.}
\end{figure}

Our predictions for the elastic electric and magnetic form factors of the charged roper resonance are shown in Figure \ref{fig:GEMR}. The electric charge radius ratio of the roper resonance to the proton is $r^R_{E}/r^p_{E} \approx 1.5$, the magnetic momentum radius ratio is also $r^R_{M}/r^p_{M} \approx 1.5$. By comparison, a Faddeev equation study predicts $r^R_{E}/r^p_{E} = 1.8$ and $r^R_{M}/r^p_{M} = 1.6$ \cite{Segovia2015}.

\section{Summary and conclusion}\label{sec:conclusion}

 In this paper, we study the elastic electric and magnetic form factors of the proton, neutron and the charged roper resonance systematically in a constituent quark model. We integrate three ingredients in our study, which are crucial for a successful description of the form factors:
\begin{enumerate}[itemindent=0em,label=\roman*)]
 \item We consider the mixing effect and use Eq. (\ref{eq:mixingPN}) as the proton/neutron state. The mixing between $|\psi_1\rangle$ and $|\psi_3\rangle$ contributes to $G_E^n$ majorly, $|\psi_4\rangle$ itself has a minor contribution.  
 \item We explore the effect of a running coupling constant. It turns out that a running coupling softens the form factor and makes the results consistent with observed values much better.
 \item We propose a running quark mass, and deduce the factorization formula of the magnetic form factor, Eq. (\ref{eq:magneticFFMQ}). The running quark mass is responsible for the decreasing of the $\mu_p G_E^p(Q^2)/G_M^p(Q^2)$ as $Q^2$ increases. Our calculation shows that $\mu_p G_E^p(Q^2)/G_M^p(Q^2) \approx M_q(Q^2/9)/M_q(0)$ upto $Q^2 \approx 4 \text{ GeV}^2$.
\end{enumerate}

Our results for the elastic electric and magnetic form factors of proton and neutron are fairly satisfactory, except that there are small deviations for $G_E^p$, $G_M^p$ and $G_M^n$ in the chiral phase transition region, and that $G_E^n$ decreases too fast as $Q^2 > 2\text{ GeV}^2$. We also predict the elastic electric and magnetic form factor of the charged roper resonance. The electric charge and magnetic momentum radius ratios of the roper resonance to the proton are $r^R_{E}/r^p_{E} \approx r^R_{M}/r^p_{M} \approx 1.5$.

\section*{Acknowledgments}
We acknowledge helpful conversations with Xian-Hui Zhong and Ming-Sheng Liu. This work is supported by: the National Natural Science Foundation of China under contracts No. 12005060.

\section*{Appendix A: The wavefuntion and classification}\label{sec:appendixA}

\setcounter{equation}{0}
\renewcommand{\theequation}{A\arabic{equation}}
\setcounter{figure}{0}
\setcounter{table}{0}
\renewcommand{\thefigure}{A\arabic{figure}}
\renewcommand{\thetable}{A\arabic{table}}

The total wave function decomposes $ \Psi = \phi \chi F \psi$, where $\phi,\, \chi,\, F,\, \psi$ are the color, spin, flavour and space wavefunctions.

The color wave function is the 1-dimensional representation of the $SU(3)_c$ group and it is totally antisymmetric under $S_3$.

\begin{equation}\label{eq:wavefunctionColor}
 \phi^A = \frac{1}{\sqrt{6}}(rgb -rbg + gbr - grb + brg - bgr),
\end{equation}
where $\{r,g,b\}$ is the fundamental representations of $SU(3)_c$. The superscript A indicates that it's antisymmetric under $S_3$ permutation. 

The direct product of three $SU(2)$ fundamental representations decompose $2\otimes 2\otimes 2 = 4 \oplus 2 \oplus 2$, so there are one 4-dimensional representation which is symmetric and two 2-dimensional representations which are mixed symmetric. Let the eigenstate with $s_z = \frac{1}{2}$ be $|\uparrow \rangle$ and that with $s_z = -\frac{1}{2}$ be $|\downarrow \rangle$. Express the spin wave function as $\chi^\sigma_{S,S_z}$, where $\sigma$ indicates the properties under $S_3$ permutation. The symmetric representation is,
\begin{eqnarray}
 \chi^S_{\frac{3}{2},\frac{3}{2}} & =& |\uparrow\uparrow\uparrow \rangle,\\
 \chi^S_{\frac{3}{2},\frac{1}{2}} & =& \frac{1}{\sqrt{3}}(|\downarrow\uparrow\uparrow \rangle + |\uparrow\downarrow\uparrow \rangle + |\uparrow\uparrow\downarrow \rangle),\\
 \chi^S_{\frac{3}{2},-\frac{1}{2}} & =& \frac{1}{\sqrt{3}}(|\downarrow\downarrow\uparrow \rangle+ |\downarrow\uparrow\downarrow \rangle + |\uparrow\downarrow\downarrow \rangle),\\
 \chi^S_{\frac{3}{2},-\frac{3}{2}} & =& |\downarrow\downarrow\downarrow \rangle.
\end{eqnarray}
The superscript S indicates that it's symmetric under $S_3$ permutation. We choose the bases of the two 2-dimensional representations to be
\begin{eqnarray}
 \chi^\rho_{\frac{1}{2},\frac{1}{2}} & =& \frac{1}{\sqrt{2}}(|\uparrow\downarrow \rangle - |\downarrow\uparrow \rangle) |\uparrow \rangle ,\\
 \chi^\rho_{\frac{1}{2},-\frac{1}{2}} & =& \frac{1}{\sqrt{2}}(|\uparrow\downarrow \rangle - |\downarrow\uparrow \rangle) |\downarrow \rangle ,\\
  \chi^\lambda_{\frac{1}{2},\frac{1}{2}} & =& \frac{1}{\sqrt{6}}(2|\uparrow\uparrow\downarrow \rangle - |\uparrow\downarrow\uparrow \rangle - |\downarrow\uparrow\uparrow \rangle) ,\\
    \chi^\lambda_{\frac{1}{2},-\frac{1}{2}} & =& \frac{1}{\sqrt{6}}(-2|\downarrow\downarrow\uparrow \rangle + |\downarrow\uparrow\downarrow \rangle + |\uparrow\downarrow\downarrow \rangle).
\end{eqnarray}
The superscript $\rho$ or $\lambda$ indicates that its symmetric or antisymmetric under the interchange $1\leftrightarrow2$.

Considering three flavours, the direct product of three $SU(3)_f$ fundamental representations decompose $3\otimes 3\otimes 3 = 10 \oplus 8 \oplus 8 \oplus 1$, so there are one 10-dimensional representation which is symmetric, two 8-dimensional representations which are mixed symmetric and one 1-dimensional representation which is antisymmetric. Let $u$, $d$, $s$ be the eigenstates with isospin and strangenese $(I_z, s) = (\frac{1}{2},0), (-\frac{1}{2}, 0), (0, -1)$. Express the flavour wave function as $F^\sigma_{I,I_z,s}$. The bases of the 10-dimensional symmetric representation form the decuplet,
\begin{equation}
 F^S = \vcenter{\includegraphics[width=0.4\textwidth]{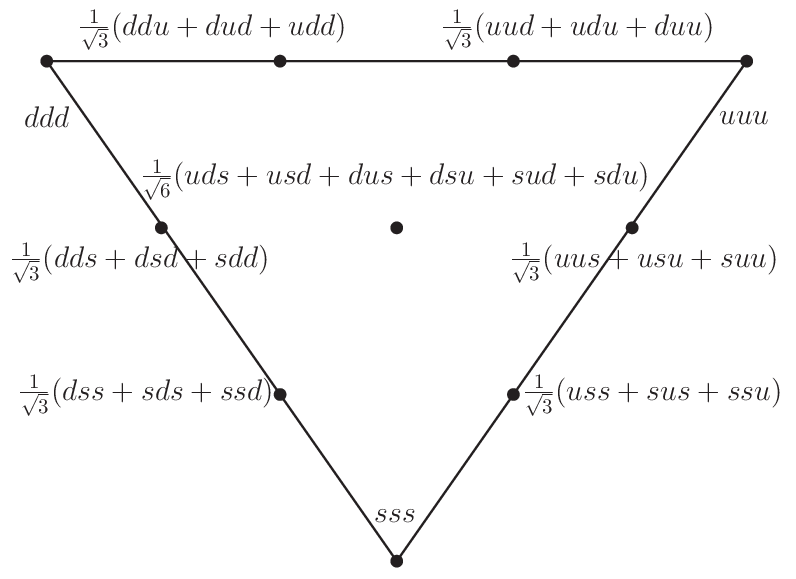}}
\end{equation}
The bases of the two 8-dimensional representations form two octets,
\begin{equation}
 F^\rho = \vcenter{\includegraphics[width=0.4\textwidth]{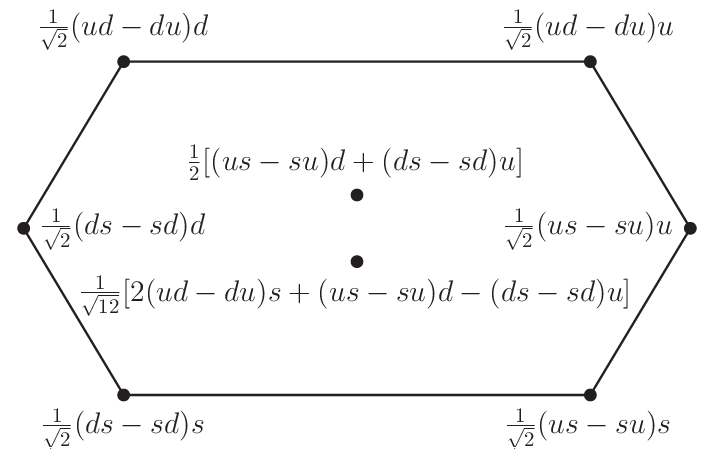}}
\end{equation}
\begin{equation}
 F^\lambda = \vcenter{\includegraphics[width=0.4\textwidth]{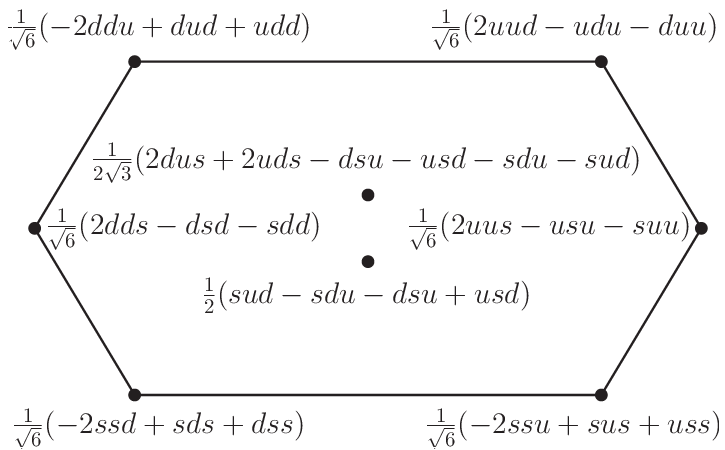}}
\end{equation}
The 1-dimensional antisymmetric representation is
\begin{equation}
 F^A = \frac{1}{\sqrt{6}}(uds -usd + dsu -dus +sud - sdu).
\end{equation}

It's an old topic to construct the three-body spacial wave function which has definite permutation properties under $S_3$ \cite{Karl1968}. Herein we use the three dimensional harmonic oscillator function of the Jacobi variables to construct the spacial wave function. The three dimensional harmonic oscillator function is 
\begin{equation}\label{eq:harmonic3D}
 u_{n l m}(\alpha r,\theta,\varphi) = R_{nl}(\alpha r)Y_{lm}(\theta,\varphi),
\end{equation}
where $Y_{lm}(\theta,\varphi)$ is the spheric harmonic function. The radial function
\begin{equation}\label{eq:harmonic3Dr}
 R_{nl}(\alpha r) = N_n (\alpha r)^l e^{-(\alpha r)^2/2}\mathcal{F}(-n, l+\frac{3}{2}, \alpha^2 r^2),
\end{equation}
where the normalization constant is $N_n = [\frac{\alpha^3 2^{l-n+2} (2l+2n+1)!!}{\sqrt{\pi}n![(2l+1)!!]^2}]^{1/2}$ and $\mathcal{F}(-n, l+\frac{3}{2}, \alpha^2 r^2)$ is the Kummer confluent hypergeometric function. In Eq. (\ref{eq:harmonic3D}) and (\ref{eq:harmonic3Dr}), $\alpha$ is a parameter related to the eigenfrequency, $\omega$, of the harmonic oscillator, i.e. $\alpha \propto \sqrt{\omega}$.

\begin{table}[b!]\centering
\caption{\label{tab:spacialWF} Spatial wave functions having definite permutation properties upto $N=2$. $\langle l_\rho l_\lambda m(M_L-m)| LM_L\rangle$ is the Clebsch-Gordan coefficient.}
\begin{tabular}{c|c|c}
\hline
$N=0$ & $\psi^S_{000}$ & $u_{000}(\bm{\rho})u_{000}(\bm{\lambda})$\\\hline
\multirow{2}{*}{$N=1$} & $\psi^\rho_{11M_L}$ & $u_{01M_L}(\bm{\rho})u_{000}(\bm{\lambda})$\\
 & $\psi^\lambda_{11M_L}$ & $u_{000}(\bm{\rho})u_{01M_L}(\bm{\lambda})$\\\hline
\multirow{7}{*}{$N=2$}  & $\psi^S_{200}$ & $ \frac{1}{\sqrt{2}}[ u_{100}(\bm{\rho})u_{000}(\bm{\lambda}) + u_{000}(\bm{\rho})u_{100}(\bm{\lambda}) ]$\\
& $\psi^\lambda_{200}$ & $ \frac{1}{\sqrt{2}}[ -u_{100}(\bm{\rho})u_{000}(\bm{\lambda}) + u_{000}(\bm{\rho})u_{100}(\bm{\lambda}) ]$\\
& $\psi^\rho_{200}$ & $  u_{01m}(\bm{\rho})u_{01-m}(\bm{\lambda}) \langle 11m-m)| 00\rangle$ \\
& $\psi^A_{21M_L}$ & $  u_{01m}(\bm{\rho})u_{01(M_L-m)}(\bm{\lambda}) \langle 11m(M_L-m)| 1M_L\rangle$\\
& $\psi^S_{22M_L}$ & $ \frac{1}{\sqrt{2}}[ u_{02M_L}(\bm{\rho})u_{000}(\bm{\lambda}) + u_{000}(\bm{\rho})u_{02M_L}(\bm{\lambda}) ]$\\
& $\psi^\lambda_{22M_L}$ & $ \frac{1}{\sqrt{2}}[ u_{02M_L}(\bm{\rho})u_{000}(\bm{\lambda}) - u_{000}(\bm{\rho})u_{02M_L}(\bm{\lambda}) ]$\\
& $\psi^\rho_{22M_L}$ & $ u_{01m}(\bm{\rho})u_{01(M_L-m)}(\bm{\lambda}) \langle 11m(M_L-m)| 2M_L\rangle$\\
\hline
\end{tabular}
\end{table}

The spacial wave function $\psi^\sigma_{NLM_L}$ could be constructed in the form 
\begin{eqnarray}\nonumber
\psi^\sigma_{NLM_L} &=&\sum\limits_{n_\rho l_\rho m_\rho n_\lambda l_\lambda m_\lambda} C^{n_\rho l_\rho m_\rho n_\lambda l_\lambda m_\lambda}\\\label{eq:spacialWF}
&& \cdot u_{n_\rho l_\rho m_\rho}(\bm{\rho}) u_{n_\lambda l_\lambda m_\lambda} (\bm{\lambda}),
\end{eqnarray}
where $N =2(n_\rho + n_\lambda)+l_\rho + l_\lambda$, $L$ is the orbital angular momentum satisfying the vector equation $\bm{L} = \bm{l_\rho} + \bm{l_\lambda}$, and $M_L = m_\rho + m_\lambda$. Spatial wave functions upto $N=2$
are listed in Table \ref{tab:spacialWF}.

\begin{table}[b!]\centering
\caption{\label{tab:SU6WF} $SU(6)$ wave functions having definite permutation properties, labelled as $N_6^\sigma = \sum\limits_{\sigma_2,\sigma_3} C^{\sigma_2,\sigma_3} N_3^{\sigma_3} \otimes N_2^{\sigma_2}$.}
\begin{tabular}{c|l}
\hline
$56^S$ & $ 10^S \otimes 4^S$, $ \frac{1}{\sqrt{2}} (8^\lambda \otimes 2^\lambda + 8^\rho \otimes 2^\rho )$ \\\hline
$70^\lambda$ & $ 10^S \otimes 2^\lambda$, $8^\lambda \otimes 4^S$, $ \frac{1}{\sqrt{2}} ( 8^\rho \otimes 2^\rho -8^\lambda \otimes 2^\lambda )$, $1^A \otimes 2^\rho$\\\hline
$70^\rho$ & $ 10^S \otimes 2^\rho$, $8^\rho \otimes 4^S$, $\frac{1}{\sqrt{2}} ( 8^\rho \otimes 2^\lambda + 8^\lambda \otimes 2^\rho )$, $-1^A \otimes 2^\lambda$\\\hline
$20^A$ & $ 1^A \otimes 4^S$, $\frac{1}{\sqrt{2}} (8^\lambda \otimes 2^\rho -8^\rho \otimes 2^\lambda )$\\
\hline
\end{tabular}
\end{table}

Combining the flavour and spin degree of freedoms, one obtain the $SU(6)$ wave function. The direct product of three $SU(6)$ fundamental representations decompose $6\otimes 6\otimes 6 = 56 \oplus 70 \oplus 70 \oplus 20$, so there are one 56-dimensional representation which is symmetric, two 70-dimensional representations which are mixed symmetric and one 20-dimensional representation which is antisymmetric. $SU(6)$ wave functions having definite permutation properties can be constructed, which are listed in Table \ref{tab:SU6WF}.

The color wave function is always the singlet, so the total wave function could be classified by $|SU(6)\otimes O(3) \rangle = |N_6, ^{2S+1}N_3,N,L,J\rangle$. The symmetric $SU(6)\otimes O(3)$ wave functions upto $N=2$ are listed in Table \ref{tab:totalWF}.

\begin{widetext}

\begin{table}[t!]\centering
\caption{\label{tab:totalWF} Symmetric $SU(6)\otimes O(3)$ wave functions upto $N=2$. $[\cdots]_J$ means combining the spin $(S,S_z)$ and orbital angular momentum $(L,M_L)$ with Clebsch-Gordan coefficients to form total angular momentum $(J,M)$.}
\begin{tabular}{c|c|c}
\hline
\multirow{2}{*}{$N=0$} & $| 56,^2 8, 0, 0, (\frac{1}{2})^+ \rangle$ & $\frac{1}{\sqrt{2}}[(F^\lambda \chi^\lambda + F^\rho \chi^\rho)\psi^S_{000}]_{\frac{1}{2}}$ \\
& $| 56,^4 10, 0, 0, (\frac{3}{2})^+ \rangle$ & $ [F^S \chi^S \psi^S_{000}]_{\frac{3}{2}}$ \\\hline
\multirow{3}{*}{$N=1$} & $| 70,^2 8, 1, 1, J^- \rangle$ & $\frac{1}{2}[(F^\rho \chi^\rho-F^\lambda \chi^\lambda )\psi^\lambda_{11M_L}+ (F^\lambda \chi^\rho+F^\rho \chi^\lambda )\psi^\rho_{11M_L}]_J$\\
& $| 70,^4 8, 1, 1, J^- \rangle$ & $\frac{1}{\sqrt{2}}[F^\lambda \chi^S\psi^\lambda_{11M_L}+ F^\rho \chi^S\psi^\rho_{11M_L}]_J$\\ 
& $| 70,^2 10, 1, 1, J^- \rangle$ & $\frac{1}{\sqrt{2}}[F^S \chi^\lambda\psi^\lambda_{11M_L}+ F^S \chi^\rho\psi^\rho_{11M_L}]_J$\\\hline
\multirow{11}{*}{$N=2$}  & $| 56,^2 8, 2, 0, (\frac{1}{2})^+ \rangle$ & $\frac{1}{\sqrt{2}}[(F^\lambda \chi^\lambda + F^\rho \chi^\rho)\psi^S_{200}]_{\frac{1}{2}}$ \\
&$| 56,^2 8, 2, 2, J^+ \rangle$ & $\frac{1}{\sqrt{2}}[(F^\lambda \chi^\lambda + F^\rho \chi^\rho)\psi^S_{22M_L}]_{J}$\\
& $| 56,^4 10, 2, 0, (\frac{3}{2})^+ \rangle$ & $ [F^S \chi^S \psi^S_{200}]_{\frac{3}{2}}$ \\
& $| 56,^4 10, 2, 2, J^+ \rangle$ & $ [F^S \chi^S \psi^S_{22M_L}]_{J}$ \\
& $| 70,^2 8, 2, 0, (\frac{1}{2})^+ \rangle$ & $\frac{1}{2}[(F^\rho \chi^\rho - F^\lambda \chi^\lambda)\psi^\lambda_{200}+ (F^\rho \chi^\lambda + F^\lambda \chi^\rho)\psi^\rho_{200}]_{\frac{1}{2}}$\\
& $| 70,^2 8, 2, 2, J^+ \rangle$ & $\frac{1}{2}[(F^\rho \chi^\rho - F^\lambda \chi^\lambda)\psi^\lambda_{22M_L}+ (F^\rho \chi^\lambda + F^\lambda \chi^\rho)\psi^\rho_{22M_L}]_{\frac{1}{2}}$\\
& $| 70,^4 8, 2, 0, (\frac{3}{2})^+ \rangle$ & $\frac{1}{\sqrt{2}}[F^\lambda \chi^S\psi^\lambda_{200}+ F^\rho \chi^S\psi^\rho_{200}]_{\frac{3}{2}}$\\
& $| 70,^4 8, 2, 2, J^+ \rangle$ & $\frac{1}{\sqrt{2}}[F^\lambda \chi^S\psi^\lambda_{22M_L}+ F^\rho \chi^S\psi^\rho_{22M_L}]_{J}$\\
& $| 70,^2 10, 2, 0, (\frac{1}{2})^+ \rangle$ & $\frac{1}{\sqrt{2}}[F^S \chi^\lambda\psi^\lambda_{200}+ F^S \chi^\rho\psi^\rho_{200}]_{\frac{1}{2}}$\\
& $| 70,^2 10, 2, 2, J^+ \rangle$ & $\frac{1}{\sqrt{2}}[F^S \chi^\lambda\psi^\lambda_{22M_L}+ F^S \chi^\rho\psi^\rho_{22M_L}]_{J}$\\
& $| 20,^2 8, 2, 1, J^+ \rangle$ & $\frac{1}{\sqrt{2}}[ (F^\lambda \chi^\rho - F^\rho \chi^\lambda)\psi^A_{21M_L}]_{J}$\\
\hline
\end{tabular}
\end{table}

\setcounter{equation}{0}
\renewcommand{\theequation}{B\arabic{equation}}
\setcounter{figure}{0}
\setcounter{table}{0}
\renewcommand{\thefigure}{B\arabic{figure}}
\renewcommand{\thetable}{B\arabic{table}}

\section*{Appendix B: The matrix element of the hamiltonian}\label{sec:appendixB}

The total wave function is of this form specifically,
\begin{equation}
 \Psi(\alpha_x) =N_\Psi\sum\limits_{\sigma} \Psi^\sigma_x =N_\Psi\sum\limits_{\sigma} \phi^A F^{\sigma3} [\chi^{\sigma2}  \psi^{\sigma s}_{LM_L n_\rho l_\rho n_\lambda l_\lambda}]_J,
\end{equation}
where $N_\Psi$ is normalization constant, $\sigma2$, $\sigma3$ and $\sigma s$ indicate the permutation properties of the spin, flavour and spacial wave functions, $[\cdots]_J$ means combining the spin $(S,S_z)$ and orbital angular momentum $(L,M_L)$ with Clebsch-Gordan coefficients to form total angular $(J,M)$.
 
The spin-spin contact interaction, 
\begin{equation}
 \langle \Psi^{\sigma'}_y | V(\rho) \bm{s}_1\cdot \bm{s}_2| \Psi^\sigma_x \rangle = \delta_{L_y,L_x}\delta_{\rho_y,\rho_x}\delta_{\lambda_y,\lambda_x} \langle n_{\rho_y}l_{\rho_y} | V(\rho) | n_{\rho_x}l_{\rho_x} \rangle \langle n_{\lambda_y}l_{\lambda_y} | n_{\lambda_x}l_{\lambda_x} \rangle  \langle \chi'||\bm{s}_1\cdot \bm{s}_2 ||\chi\rangle \frac{1}{\sqrt{2S_y +1}}.
\end{equation}
The reduced matrix element
\begin{equation}
\langle \chi'||\bm{s}_1\cdot \bm{s}_2 ||\chi\rangle = \begin{pmatrix} \frac{1}{2}&0&0 \\ 0& -\frac{3\sqrt{2}}{4}&0 \\ 0&0&\frac{\sqrt{2}}{4} \end{pmatrix},
\end{equation}
with the columns and rows corresponding $\{\chi',\chi\} = \{ \chi^S, \chi^\rho, \chi^\lambda \}$.

Let $\bm{S}_{12} = \bm{s}_1 + \bm{s}_2$, $\bm{L}_\rho = \bm{L}_{12}$. The spin-orbital interaction
\begin{equation}
 \langle \Psi^{\sigma'}_y | V(\rho) \bm{S}_{12}\cdot \bm{L}_\rho| \Psi^\sigma_x \rangle = (-1)^{S_y + L_x + J}\begin{Bmatrix}
 L_y & L_x & 1\\
 S_x & S_y & J
\end{Bmatrix}
 \langle \chi' ||\bm{S}_{12}|| \chi \rangle
 \langle L_y n_{\rho_y}l_{\rho_y}n_{\lambda_y}l_{\lambda_y} ||V(\rho)\bm{L}_\rho|| L_x n_{\rho_x}l_{\rho_x}n_{\lambda_x}l_{\lambda_x}\rangle,
 \end{equation}
where $\begin{Bmatrix}
 L_y & L_x & 1\\
 S_x & S_y & J
\end{Bmatrix}$ is the 6j symbol. The spin reduced matrix is
\begin{equation}
\langle \chi' ||\bm{S}_{12}|| \chi \rangle=
 \begin{pmatrix}
  2\sqrt{\frac{5}{3}} & 0 & \frac{2}{\sqrt{3}}\\
  0 & 0 & 0\\
  -\frac{2}{\sqrt{3}} & 0 & 2\sqrt{\frac{2}{3}}
 \end{pmatrix}.
\end{equation}
The spacial reduced matrix is further reduced by using Wigner-Eckart theory again,
\begin{eqnarray}\nonumber
 \langle L_y n_{\rho_y}l_{\rho_y}n_{\lambda_y}l_{\lambda_y} ||V(\rho)\bm{L}_\rho|| L_x n_{\rho_x}l_{\rho_x}n_{\lambda_x}l_{\lambda_x}\rangle &=& \delta_{l_{\lambda_x},l_{\lambda_y}} (-1)^{l_{\rho_y} + l_{\lambda_x} + L_x +1} \begin{Bmatrix}
 l_{\rho_y} & l_{\rho_x} & 1\\
 L_x & L_y & l_{\lambda_y}
 \end{Bmatrix}
 \langle n_{\rho_y} l_{\rho_y} |V(\rho)| n_{\rho_x}l_{\rho_x}\rangle \\
 && \cdot \langle n_{\lambda_y} l_{\lambda_y} | n_{\lambda_x}l_{\lambda_x}\rangle
  \sqrt{(2L_x+1)(2L_y+1)}\langle l_{\rho_y}||\bm{L}_\rho||l_{\rho_x} \rangle,
\end{eqnarray}
where $\langle l_{\rho_y}||\bm{L}_\rho||l_{\rho_x} \rangle = \delta_{l_{\rho_x},l_{\rho_y}} \sqrt{l_{\rho_y}(l_{\rho_y}+1)(2l_{\rho_y}+1)}$.

The tensor operator could be written as $T = 3\bm{s}_1\cdot \hat{\bm{\rho}} \bm{s}_2\cdot \hat{\bm{\rho}} - \bm{s}_1\cdot\bm{s}_2 = \bm{R}_2 \cdot \bm{S}_2$, where $\hat{\bm{\rho}} = \bm{\rho}/\rho$. $\bm{R}_2$ and $\bm{S}_2$ are second order irreducible tensor,
$$\bm{R}_2^\mu = \begin{pmatrix}
\frac{\sqrt{3}}{2}\hat{\rho}_+^2 \\
-\sqrt{3}\hat{\rho}_+ \hat{\rho}_z\\
\frac{1}{\sqrt{2}}(3\hat{\rho}_z^2 -1)\\
\sqrt{3}\hat{\rho}_- \hat{\rho}_z\\
\frac{\sqrt{3}}{2}\hat{\rho}_-^2 \\
\end{pmatrix}
\text{ and }
\bm{S}_2^\mu = \begin{pmatrix}
\frac{\sqrt{3}}{2} s_{1+}s_{2+}\\
-\frac{\sqrt{3}}{2}(s_{1+}s_{2z} + s_{1z}s_{2+})\\
\frac{1}{\sqrt{2}} (2s_{1z}s_{2z} - \frac{s_{1+}s_{2-} + s_{1-}s_{2+}}{2}) \\
\frac{\sqrt{3}}{2}(s_{1-}s_{2z} + s_{1z}s_{2-})\\
\frac{\sqrt{3}}{2} s_{1-}s_{2-}
\end{pmatrix},
$$
where $\hat{\rho}_{\pm} = \hat{\rho}_x \pm i\hat{\rho}_y$ and $s_{i\pm} = s_{ix} \pm is_{iy}$. The tensor interaction
\begin{equation}
 \langle \Psi^{\sigma'}_y | V(\rho) \bm{R}_2 \cdot \bm{S}_2 | \Psi^\sigma_x \rangle = (-1)^{S_y + L_x + J}\begin{Bmatrix}
 L_y & L_x & 2\\
 S_x & S_y & J
\end{Bmatrix}
 \langle \chi' ||\bm{S}_{2}|| \chi \rangle
 \langle L_y n_{\rho_y}l_{\rho_y}n_{\lambda_y}l_{\lambda_y} ||V(\rho) \bm{R}_2|| L_x n_{\rho_x}l_{\rho_x}n_{\lambda_x}l_{\lambda_x}\rangle.
 \end{equation}
The spin reduced matrix is
\begin{equation}
\langle \chi' ||\bm{S}_{2}|| \chi \rangle=
 \begin{pmatrix}
  \sqrt{\frac{5}{2}} & 0 & \sqrt{\frac{5}{2}}\\
  0 & 0 & 0\\
  -\sqrt{\frac{5}{2}} & 0 & 0
 \end{pmatrix}.
\end{equation}
The spacial reduced matrix is further reduced by using Wigner-Eckart theory again,
\begin{eqnarray}\nonumber
 \langle L_y n_{\rho_y}l_{\rho_y}n_{\lambda_y}l_{\lambda_y} ||V(\rho) \bm{R}_2|| L_x n_{\rho_x}l_{\rho_x}n_{\lambda_x}l_{\lambda_x}\rangle &=& \delta_{l_{\lambda_x},l_{\lambda_y}} (-1)^{l_{\rho_y} + l_{\lambda_x} + L_x +2} \begin{Bmatrix}
 l_{\rho_y} & l_{\rho_x} & 2\\
 L_x & L_y & l_{\lambda_y}
 \end{Bmatrix}
 \langle n_{\rho_y} l_{\rho_y} |V(\rho)| n_{\rho_x}l_{\rho_x}\rangle \\
 && \cdot \langle n_{\lambda_y} l_{\lambda_y} | n_{\lambda_x}l_{\lambda_x}\rangle
  \sqrt{(2L_x+1)(2L_y+1)}\langle l_{\rho_y}|| \bm{R}_2||l_{\rho_x} \rangle,
\end{eqnarray}
where $\langle l_{\rho_y}|| \bm{R}_2||l_{\rho_x} \rangle = \langle l_{\rho_x }200|l_{\rho_y} 0 \rangle \sqrt{2(2l_{\rho_x}+1)}$.
\end{widetext}

\bibliographystyle{unsrt}
\bibliography{../references/PFF}

\end{document}